\documentclass[preprint]{aastex}

\slugcomment{}

\shorttitle{Infrared Parallaxes}
\shortauthors{Vrba et al.}

\begin{document}

\title{Preliminary Parallaxes of 40 L and T Dwarfs from the \\ U.S. Naval Observatory Infrared Astrometry
Program}

\author{F. J. Vrba, A. A. Henden\altaffilmark{1}, C. B. Luginbuhl, H. H. Guetter,
J. A. Munn, and B. Canzian}
\affil{U.S. Naval Observatory, Flagstaff Station, P.O. Box 1149, Flagstaff, AZ
    86002--1149}
\affil{fjv@nofs.navy.mil, aah@nofs.navy.mil, cbl@nofs.navy.mil, guetter@nofs.navy.mil,
jam@nofs.navy.mil, blaise@nofs.navy.mil}

\author{A. J. Burgasser\altaffilmark{2}}
\affil{University of California, Los Angeles, Division of Astronomy and Astrophysics, \\ 405 Hilgard Ave.,
Los Angeles, CA 90095--1562}
\affil{adam@astro.ucla.edu}

\author{J. Davy Kirkpatrick}
\affil{California Institute of Technology, IPAC, \\ 770 S. Wilson Ave., MS 100--22, Pasadena, CA 91125}
\affil{davy@ipac.caltech.edu}

\author{X. Fan}
\affil{Steward Observatory, The University of Arizona, 933 N. Cherry Ave., Tucson, AZ 85721}
\affil{fan@as.arizona.edu}

\author{T. R. Geballe}
\affil{Gemini Observatory, 670 N. A'ohoku Place, Hilo, HI 96720}
\affil{tgeballe@gemini.edu}

\author{D. A. Golimowski}
\affil{Department of Physics and Astronomy, The Johns Hopkins University \\
3400 N. Charles St., Baltimore, MD 21218}
\affil{dag@pha.jhu.edu}

\author{G. R. Knapp}
\affil{Department of Astrophysical Sciences, Princeton University, Princeton, NJ 08544}
\affil{gk@astro.princeton.edu}

\author{S. K. Leggett}
\affil{United Kingdom Infrared Telescope, Joint Astronomy Centre, \\
660 N. A'ohoku Place, Hilo, HI 96720}
\affil{s.leggett@jach.hawaii.edu}

\author{D. P. Schneider}
\affil{Pennsylvania State University,  Department of Astronomy and Astrophysics, \\
525 Davey Lab., University Park, PA 16802}
\affil{dps@astro.psu.edu}

\and

\author{J. Brinkmann}
\affil{Apache Point Observatory, P.O. Box 59, Sunspot, NM 88349-0059}
\affil{jb@apo.nmsu.edu}

\altaffiltext{1}{Universities Space Research Association}
\altaffiltext{2}{Hubble Fellow}

\begin{abstract} 

We present preliminary trigonometric parallaxes and proper motions for 22 L dwarfs and 18 T 
dwarfs measured using the ASTROCAM infrared imager on the U.S. Naval Observatory (USNO) 1.55--m
Strand Astrometric reflector. The results presented 
here are based on observations obtained between September 2000 and November 2002;
about half of the objects have an observational time baseline of $\Delta$t~=~1.3 yr and
half $\Delta$t~=~2.0 yr. Despite these short time baselines, the astrometric 
quality is sufficient to produce significant new results, especially for the nearer T dwarfs.
Seven objects are in common with the USNO optical CCD parallax program for quality control
and seven in common with the ESO 3.5--m NTT parallax program. We compare astrometric quality with both of 
these programs. Relative to absolute parallax corrections are made by employing 2MASS and/or SDSS photometry
for reference frame stars. We combine USNO infrared and optical parallaxes with the best available CIT 
system photometry to determine $M_J$, $M_H$, and $M_K$ values for 37 L dwarfs between spectral types 
L0 to L8 and 19 T dwarfs between spectral types T0.5 and T8 and present selected absolute magnitude versus 
spectral type and color diagrams, based on these results. Luminosities and temperatures are estimated 
for these objects. Of special interest are the  distances of several objects which
are at or near the L--T dwarf boundary so that this important transition can be better understood.
The previously reported early-mid T dwarf luminosity excess is clearly confirmed and found to
be present at J, H, and K. The large number of objects that populate this luminosity excess region indicates
that it cannot be due entirely to selection effects. The T dwarf sequence is extended to $M_J$~$\approx$~16.9
by 2MASS J041519$-$0935 which, at d~=~5.74 pc, is found to be the least luminous [log(L/L$_{\odot}$)~=~-5.58]
and coldest (T$_{\rm eff}$~$\approx$~760 K ) brown dwarf known. Combining results from this paper
with earlier USNO CCD results we find that, in contrast to the L dwarfs, there are no examples of low velocity 
(V$_{tan}$$<$ 20 km $s^{-1}$) T dwarfs. This is consistent with the T dwarfs in this study being generally older 
than the L dwarfs. We briefly discuss future directions for the USNO infrared astrometry program.

\end{abstract}


\section{Introduction}

After the era of photographic proper motion surveys (e.g. Luyten 1979) revealed late M stars
close to the limit of stellar hydrogen burning, a long, and largely frustrating, pursuit of
sub--stellar objects was begun by many researchers. These efforts were motivated
by the overarching desire to understand the Galactic mass and luminosity distributions of
putative objects that would bridge the gap between the lowest mass stars and giant planets
and by the fact that no theory of star formation could be considered complete without accounting
for the mass function of such objects. Becklin \& Zuckerman (1988) identified GD 165B
as the first object clearly cooler than an M dwarf, followed several years later
by the discovery of the first `methane dwarf', Gliese 229B \citep{nak95,opp95}; an object cold enough
that its spectrum shows strong methane absorption, similar to the giant gas planets. 
These `brown dwarfs' became the prototypes for L dwarfs \citep{kir99,mar99} and T dwarfs
\citep{bur02a,geb02}, respectively. 

It was not, however, until deep, large--scale optical surveys (the Sloan Digital Sky Survey [SDSS]\footnote{\bf
www.sdss.org}; \citealp{york00,abazajian03}) and 
near--infrared surveys (the Two Micron All Sky Survey [2MASS\footnote{\bf www.ipac.caltech.edu/2mass}; 
Skrutskie et al. 1997] and the Deep Near Infrared Survey of the Southern Sky [DENIS\footnote{\bf
cdsweb.u-strasbg.fr/denis.html}; Delfosse et al. 1997, Epchtein 1997]) of the sky were undertaken that significant 
numbers of field brown dwarfs were revealed. L dwarfs from these surveys have been identified by many
authors \citep{del97,kir99,kir00,fan00,haw02,geb02,sch02}, as have T dwarfs
\citep{bur99,str99,tsv00,leg00,bur02a,geb02,bur03a,kna04}, amongst many others. Comprehensive summaries of field brown dwarf
discoveries are maintained at the web sites maintained by Kirkpatrick for L dwarfs
({\bf spider.ipac.caltech.edu/staff/davy/ARCHIVE}), by Burgasser for T dwarfs 
({\bf www.astro.ucla.edu/adam/homepage/research/tdwarf}), and by Leggett for both L and T dwarfs
({\bf www.jach.hawaii.edu/skl/LTdata.html}). Currently there are more than 250
L dwarfs and 58 T dwarfs known.

Unlike stars, brown dwarfs are not massive enough to sustain continuous hydrogen fusion in their
cores, but cool continually from their birth. Somewhere between early and mid--L is 
the crossover between hydrogen--burning stars and brown dwarfs. Unfortunately, other than objects 
in clusters \citep{bas00}, it is difficult to establish ages for brown dwarfs, since their spectra do not always exhibit a
known direct indicator of age such as from Li destruction. This results in degeneracies amongst mass, age, and luminosity. 
However, for all but the youngest objects, brown dwarf radii are largely independent of mass and age, and
all are similar to the radius of Jupiter \citep{cha00}. Thus luminosity scales well with T$_{\rm eff}^4$, with L dwarfs having 
surface temperatures in the range 2200--1400 K, while T dwarfs have temperatures down 
to about 700~K (e.g. Kirkpatrick et al. 2000; Leggett et al. 2001; Leggett et al. 2002; Dahn et al. 2002;
Burgasser at al. 2002a; Golimowski et al. 2004; this paper). Obviously, an accurate measurement of the distances to 
these objects is required to determine their luminosities and temperatures, along with understanding many other
issues such as their cooling curves and surface flux redistribution due to atmospheric dust formation. 

In earlier work, U.S. Naval Observatory (USNO) optical CCD parallaxes and proper motions were presented 
for eight late M dwarfs, 
17 L dwarfs, and three T dwarfs (Dahn et al. 2002, hereafter D02). Most recently Tinney, Burgasser, 
\& Kirkpatrick (2003, hereafter TBK03) presented 
near--infrared parallaxes and proper motions of 9 T dwarfs. In this paper we present preliminary 
trigonometric parallaxes and proper motions, obtained at near--infrared wavelengths, of 22 L dwarfs and 18 T 
dwarfs plus four additional L or T dwarf companion objects in binaries. 
We feel compelled to present preliminary parallaxes and proper motions now due to the intense interest
by the  community in distance determinations to these objects, rather than waiting
approximately another two years of observational time baseline before final results would be
available for most of the objects. Final results will be presented in later papers.

\section{Development of Near--Infrared Astrometric Capabilities at USNO}

In the mid-1990s USNO anticipated a need to have the capability of carrying out high accuracy
relative astrometry at near--infrared wavelengths. This was based on the possibility that results
from the upcoming 2MASS, DENIS, and SDSS sky surveys might reveal large numbers of brown dwarfs
and other cool and dust-embedded objects which would be better detected in the near--infrared than
at visible wavelengths. This need was also consistent with the USNO mission of testing the
astrometric capabilities of new technology array detectors and extending this investigation to longer
wavelengths. While the near--infrared offered the prospects of better detection for cool objects with
smaller differential color refraction and somewhat better seeing than in the optical, the effects 
on astrometry of telescope and variable sky background radiation, along with the additional infrared camera optics needed to 
apodize emissive telescope parts and the performance of infrared array detectors themselves, were not known. 
Astrometric testing was carried out between 1995 and 1999 at USNO using a Rockwell HgCdTe 256$^2$ 
(NICMOS-II) array in a camera not optimized for astrometric work. Repeated observations were obtained
during this time frame of stars in the clusters M67 and NGC~7790 with J,H,K magnitudes between about 
11 and 14. For observations with seeing $\le$~1.5 arcsec, we found mean errors of unit weight 
for a single observation of 7, 10, and 11 milli--arcseconds (mas) for J, H, and K, respectively \citep{vrb00}. These
results, while not as good as those obtained with CCDs \citep{dah97}, encouraged us to pursue 
instrumentation specifically designed for carrying out a routine astrometric program in the near--infrared.

A major problem encountered was that, at reasonable pixelization, 256$^2$ format devices offer fields of view
which typically are not large enough to present an adequate reference frame. Thus, in 1993 USNO
joined with partner institution the National Optical Astronomical Observatories to fund the
development of the ALADDIN 1024$^2$ InSb array detector \citep{fow96} at Santa Barbara Research Corporation (now Raytheon
Vision Systems). This partnership led to the successful development of ALADDIN
arrays, which are now used throughout astronomy. In 1996 USNO and the Naval Research Laboratory
jointly began designing an ALADDIN--based camera. Mauna Kea Infrared (MKI) was contracted to build this camera,
known as ASTROCAM, which was delivered to the USNO, Flagstaff Station in August 1999. For 20 months
ASTROCAM was operated with an engineering grade detector, during which time we optimized operation of the 
camera and developed an operational plan for making brown dwarf astrometric observations. 
In April 2000 a science grade ALADDIN array was installed, and in September 2000 the 
full parallax and proper motion program for brown dwarfs was initiated. 

\section{Instrumentation}

ALADDIN array detectors \citep{fow96} have 27~$\mu$m pixels with essentially 100\% fill factor and approximately 
85\% quantum efficiency between 0.9 and 5 $\mu$m. The array is designed with quadrant architecture and
eight outputs per quadrant. The particular array which we employ in ASTROCAM has dark current of about 
0.7 e$^-$ sec$^{-1}$ at its operating temperature of 30~K and has a full well capacity of about 2.1~x~10$^5$ e$^-$ at
its operating $-$0.8~volt bias level. With double correlated sampling we obtain a read noise of about 30 e$^-$ RMS.
Except for an avoided region near one edge, which was over-thinned during fabrication and which covers 
about 3\% of the area of the array, our device has pixel operability of $>$~99.99\%.

ASTROCAM was delivered by MKI as a turnkey system including DSP--based electronics and a graphical
user interface and is used exclusively at the 1.55-m Kaj Strand Astrometric Reflector at the USNO, Flagstaff
Station. Since one of ASTROCAM's missions is astrometric measurements, its structure was designed 
to minimize flexure. The heart of ASTROCAM is an all--reflective Offner 1:1 re-imaging system,
which eliminates refractive optics except for the entrance window and filters, which are tilted at
5$^{\circ}$ to avoid production of ghost images. Thus the ALADDIN 27~$\mu$m
pixels are at the natural telescope scale providing 0.3654 arcsec pixelization and a field of view of
about 6.2 x 6.2 arcmin. Determination of the pixel scale was accomplished by observation of stellar
clusters with well-determined astrometric positions. 
Pupil plane apodization of the telescope structure is accomplished via a mask
deposited on the secondary mirror of the Offner re-imager for the outside edge of the optics and a
light trap cone to mask the telescope primary mirror central hole. The telescope secondary support struts are not
apodized. The system is telecentric so that mis--focus does not change the focal plane scale. The Offner 
system provides image spot sizes of less than 0.5 pixel even at the very corners of the array. Field
distortion has been measured, by observation with ASTROCAM of astrometric calibration fields made with 
the Flagstaff Astrometric Scanning Transit Telescope \citep[FASTT;][]{sto97}, at less than 100 mas RMS
over the entire field, a number which is dominated by the FASTT astrometric accuracy. Two 10-position concentric
filter wheels currently house seven broadband filters (Z, J, H, K, K$^{\prime}$, K-long, L$^{\prime}$) and nine narrow-band 
astrophysical filters, plus a cold blank position. An extensive description of the ASTROCAM system is given by
Fischer et al. (2003).

\section{Observational Procedures}

ASTROCAM is scheduled on the 1.55-m telescope from 11 to 14 nights each lunation during bright time.
Approximately 75\% of the scheduled time is used for observations on the infrared parallax and
proper motion program. Observations are carried out under seeing conditions up to 2.5 arcsec FWHM, although, in practice, 
the exposure times get prohibitively long for some of the fields at 2.5 arcsec. 
Despite the fact that differential color refraction (DCR) in the near--infrared is smaller than in the optical, 
we continue the CCD parallax program practice of only obtaining observations as objects cross the 
meridian. Because of this, we do not apply DCR corrections, although we may consider testing DCR corrections on some 
fields in the future.

Prior to starting our parallax program we carried out test astrometric observations of several L and T dwarfs in the J, H, and 
K bands. Best results were obtained when L dwarfs were observed in H band and T dwarfs in J band, reflecting the highest
signal--to--background noise ratio for each kind of object. With the exception of two objects,
noted below, we continue to observe T dwarfs in J band and L dwarfs in H band.

Exposure times are set by the desire to not saturate either the parallax object or reference frame stars
and range from 30 to 60 seconds, with the number of coadds under nominal conditions ranging from 8-20,
with approximately twice as many coadds employed under the most marginal conditions. A typical integration
time for a single telescope dithered position is 10 minutes in nominal conditions and 20 minutes for a faint
field under poor conditions. These relatively long exposure times are necessary because the 1.55--m 
telescope has a small effective aperture due to its large secondary mirror.
Since our total integration times for a given field can range from 20 to 60 minutes, all exposures are guided by an
optical wavelength leaky guider. These relatively long exposure times are another reason why we do not obtain observations off the
meridian, since the large DCR between the infrared imaging and optical guiding would be transfered
to the infrared images.  

Three dithered integrations are obtained for each observation. The dither offset between positions is
10 arcsec, with the dither pattern being north--south or east-west around the nominal position, depending
on the distribution of the reference frame stars. The nominal registration of each field is repeated
to within a few pixels for each visitation of a field. We have chosen to stay with a minimum number of dithers based on our
tests which show that astrometric quality is not affected by pixel sampling even under the best seeing
conditions we experience.

\section{Data Processing and Astrometric Reduction Procedures}

The data processing procedures are again a product of extensive
testing of astrometric solutions with ASTROCAM during the period before the astrometric program
began. The first step is that all frames used in processing, program and flats, are linearized via
a process similar to that described by Luginbuhl et al. (1998) for our earlier NICMOS-II device.
Linearization improves the astrometric centroiding somewhat since it has the effect of slightly sharpening
the image profile. After linearization all program frames are flat--fielded. We use dome flats, which are composed
of the difference frames of three dome flat screen illuminated and unilluminated sets. The difference
frames are median combined into a single flat--field frame for each filter employed. After flat--fielding, the 
program frames are passed through a min/max value filter to contruct a sky frame which
is subtracted from each program frame.

Astrometry is performed on individual program frames, processed as described above, not on combined
frames. Centroiding for the parallax target object and the reference frame stars is accomplished via
a two dimensional Gaussian fit technique \citep{mon83}, which is also employed in the USNO optical CCD 
parallax program. The astrometric solutions are determined via the techniques and software developed
by D. G. Monet \citep{mon83,mon92} which are again borrowed from the USNO optical program and
modified for use in our infrared program to accomodate such factors as highly varying net backgrounds.
As in the optical CCD program, we use only linear frame constants and allow frame scale and rotation to be free parameters. 
(In fact, we do solve for second and third order frame constants, but discard them as they are always trivial
compared to their solution errors.) All reference stars used in the reference frame are given unit
weighting.

\section{What is Meant by `Preliminary' Astrometric Results?}

There are several considerations which make the parallax and proper motion results we present in this
paper `preliminary' rather than final. While our observing program is continuing, due to the intense
interest in brown dwarfs, we decided to use the observations in hand at the end of the November 2002 observing run
to provide these preliminary astrometric results. Since observations were initiated in September 2000, this
meant that about half of the fields had been observed for a time baseline $\Delta$t of about 2.0 years and about
half for only 1.3 years. Normally we would allow a minimum three years of observing to allow uniform
coverage of the observable part of the parallactic ellipse, add time baseline to the proper motion 
determinations, add to $\sqrt{n}$ statistics, and have a complete separation of parallax from proper 
motion solutions. 

Second, the parallax results we report here are only for the X (right ascension) solutions and
do not include the Y (declination) solutions weighted by error. The astrometric errors we find for
ASTROCAM are nearly as good in Y as they are in X (see \S 11) and, of course, we have to
employ both X and Y astrometry to derive proper motions. However, since the parallactic ellipse is
a circle at the ecliptic pole and a line in the ecliptic plane, the Y parallax determination is
always worse than the X determination except at the pole. Thus, Y parallax solutions with the
short $\Delta$t as of November 2002 range from having internal S/N of 10 down to 1. Rather than publishing
combined values for only those cases for which the Y determination helped the solution, we feel
that it is more straightforward to publish the X--determined solutions at this time. 
When we publish final parallaxes we will present fully combined X and Y solutions.

Third, we are not employing the best astrometric reference frames for our preliminary solutions.
The reference frames we use here are based on those stars for which we have adequate optical and/or
infrared photometry to determine a photometric parallax for use in determining the correction from
relative to absolute parallax for each frame (see \S 9 on absolute parallax corrections). Naturally, 
it is best to use as many stars as possible which are well distributed over the field to produce the most 
robust reference frame. For most fields we have at least some stars which serve as otherwise excellent 
reference frame stars, but for which we do not have adequate photometry, and thus they cannot be employed
in our solution. Although it is the case that the corrections from relative to absolute parallax are typically
dwarfed by the parallaxes of these objects, it is not formally correct to include reference stars for which we
have no photometric parallaxes. It is one of our goals to obtain the necessary photometry in order to
employ the best possible reference frames when we publish final parallaxes.

Finally, we have not solved for nor culled any stars in the reference frames which might have measurable
proper motions. This should have no effect on the parallax solutions, since parallax and proper motion
solutions are largely orthogonal and should have minimal effect on proper motions since there are numerous
reference frame stars used in each solution. Mostly this will have the effect of increasing the apparent 
errors of our global frame solutions (see \S 11 on astrometric quality).

Despite these shortcuts, the preliminary parallax and proper motions we present here  
agree with the previously published results for objects in common
and produce relatively tight spectral type -- absolute magnitude diagrams. We are confident that our
results are significant to within the errors we publish. Nonetheless, we caution that these are preliminary
results.

\section{Objects on the Program}

In Table~1 we list the full designations of the 40 L and T dwarfs or multiple systems with which we began our initial 
near--infrared parallax program in September 2000 and for all of which we are reporting preliminary
results. Hereafter, we use an abbreviated
designation for each object.  There are, in fact, 44 L and T dwarfs for which we have parallactic 
information as four of the objects (2MASS~J085035+1057AB, 2MASS~J122554$-$2739AB, 2MASS~J172811+3948AB, and
2MASS~J210115+1756AB) are known doubles, which
we discuss later. The 40 initial objects were selected primarily to complement those  which
were already being observed on the USNO optical parallax program. Specifically, this meant an
emphasis on T dwarfs and a selection of late L dwarfs, all of which were difficult to observe with
optical CCDs. Twenty two of the objects are L dwarfs and 18 are T dwarfs. 
Seven objects (2MASS J055919$-$1404, 2MASS J082519+2115, 2MASS J085035+1057, SDSS J125453$-$0122, 
SDSS J162414+0029, 2MASS J163229+1904, and 2MASS J222443$-$0158) were chosen to be observed in parallel with the 
USNO optical program in order to provide independent determinations of parallaxes and proper motions
and to serve as quality control for both programs. Some of the objects had already been published 
at the time we began our program, while others were provided to us by members of the 2MASS and SDSS
teams as part of the USNO collaboration in these surveys. Twenty two of the objects are from the
2MASS survey, while 18 are from the SDSS survey. Column two of Table~1 provides the adopted spectral type from the 
literature (see \S 13).

Column 3 of Table~1 gives the broadband filter (J or H) in which the astrometry is done for these objects.
Note that, contrary to our methodology described above (that we observe L dwarfs in the H filter and T dwarfs
in the J filter), two of the objects (SDSS J015141+1244 and SDSS J020742+0000) are T dwarfs being observed in the 
H filter. This is due to the fact that at the time we started observations they were thought to be L dwarfs,
but were later determined to be T dwarfs. Rather than starting a new series of observations we decided to
simply continue the observations in the H filter, although at the expense of optimal detection S/N.

Columns 4 and 5 give the number of nights on which each object has been observed and the timespan over
which the observations have been taken, respectively. The 18 objects between right ascension 10$^h$ and 18$^h$
have been observed for an average of 21.2 nights and an average timespan of 1.34 years, while those between
right ascension 18$^h$ and 10$^h$ have been observed for an average of 26.7 nights but with a significantly 
longer average elapsed time of 2.03 years. This is an artifact of our decision to use data taken through 
the observing run of November 2002 as the preliminary database.

Column 6 gives the mean epoch of the observations for each object.
Column 7 gives the number of reference frame stars employed in the solution,  while 
column 8 gives a code describing how the conversion from relative to absolute parallax was determined for each
reference frame, as explained in \S 9.

\section{Reference Frame Selection}

Registration of the parallax fields was determined by taking several test exposures and choosing a registration
that left as many potential reference frame stars in the field of view of the array, but left the parallax 
target object as close to the field center as possible. Once observations were begun, a few processed frames
with the best seeing available were inspected to rule out any extended objects or double stars as potential
reference frame members.  When adequate time had elapsed so that running astrometric solutions
were practical, solutions were started using the full potential reference frame and stars identified as degrading
the solution were discarded. The stars removed were typically significantly fainter than the mean reference 
frame brightness or very near the edge of the array. Finally, stars that did not have adequate photometry
from which a photometric parallax could be determined were removed, as discussed above. Since we consider the reference frames
used for these preliminary results to be provisional, we will defer, until such time as we publish final
parallax results, a full discussion of the reference frames employed, including identification and magnitude range.

\section{Relative to Absolute Parallax Corrections}

Although the target objects of this program are nearby, we nevertheless determined 
corrections from relative to absolute parallaxes via photometric parallaxes for the reference
frame stars using optical photometry from SDSS and/or infrared photometry from 2MASS.
In all cases, photometric parallaxes were derived assuming that reference frame stars are main--sequence dwarfs.
We used dereddened (using the extinction maps of Schlegel, Finkbeiner, \& Davis 1998)
SDSS colors for those reference frame stars with SDSS photometry (2MASS J121711$-$0311, 
2MASS J123739+6526, 2MASS J171145+2232, and all SDSS fields except SDSS J042348$-$0414 and SDSS J053952$-$0059, which lie 
outside the official SDSS coverage). Since the average reference frame star distance is $\approx$ 630 pc (1.59 mas,
see below) we applied the full thin disk reddening of the Schlegel, Finkbeiner, \& Davis (1998) maps.
For red stars ($i-z > 0.5$) we used the $M_{i}$ vs. $i-z$ calibration 
of Hawley et al. (2002). For blue stars ($i-z < 0.5$) we used the $M_R$ vs. $R-I$ calibration
of Siegel et al. (2002).  $r-i$ colors were transformed to $R-I$ colors using the transformations given 
by Smith et al. (2002). $r$ magnitudes were transformed to $R$ magnitudes using the equation 
$R = r - 0.21 (r-i) - 0.17$ (D. Tucker, private communication).  Only those blue stars in the color range
$0.3 < R-I < 1.5$ had photometric parallaxes derived.  This is slightly bluer than the range adopted by 
Siegel et al. (who used a blue cutoff of 0.4); the calibration still fits the dwarf sequence to $R-I = 0.3$, 
though confusion with turnoff stars becomes greater.  

For fields without SDSS photometry, but with 2MASS photometry, we transformed the 2MASS $J-H$ and $H-K$
colors to California Institute of Technology (CIT) colors \citep{eli82} using the transformations of 
Carpenter~(2001). The transformed colors were used
to estimate spectral types from the spectral type versus infrared color calibration  (transformed to the CIT system)
of Bessell \& Brett (1988). Absolute V magnitudes as a function of spectral type were taken from 
Schmidt-Kaler (1982) and converted to $M_J$ and $M_K$ via the Bessell \& Brett (1988) calibrations. Distances were
determined by averaging $(m-M)_J$ and $(m-M)_K$. Since extinction in the $J$-- and $K$--bands is only a small
fraction of that in the optical, reddening in the infrared was ignored. For the fields for which both SDSS and 2MASS 
colors were available, the average distance was taken. Column~7 of Table~1 gives the codes H, S, or IR
indicating whether the Hawley et al. (2002), Siegel et al. (2002), and/or infrared calibrations were employed.

Since we give each reference frame star equal weight in the astrometric solution, the distances in milli--arcseconds
(mas) for each star are simply averaged and a standard deviation of the mean calculated for each reference frame.
The reference frame distances in mas are added to the relative parallax astrometric solutions in mas and 
the errors of the reference frame distance corrections are added in quadrature to the relative parallax astrometric
errors. The average correction to absolute parallax is 1.59 mas, with average uncertainty 0.31 mas, and 
scatter 0.46 mas (std dev). The average correction for the 16 fields observed in the J filter is 1.70 mas, while 
for the 24 fields observed in the H filter it is 1.52 mas.

The mean ratio of $\Delta\pi_{rel\rightarrow abs}$ / $\pi_{abs}$ is
0.036 $\pm$ 0.026 indicating that the mean correction adds only a few percent to the distance of
these objects. The mean ratio of $\sigma(\Delta\pi_{rel\rightarrow abs})$ / $\sigma$($\pi_{rel}$) is
0.087 $\pm$ 0.059, indicating that the error of the corrections to absolute parallax, when added
in quadrature, adds almost nothing to the total parallax error. The mean ratio of
$\sigma(\Delta\pi_{rel\rightarrow abs})$ / $\pi_{abs}$ is 0.008 $\pm$ 0.014, confirming that
the error of the relative to absolute parallax correction is small compared to the parallaxes themselves
for these objects.

Finally, we note that there are 76 reference frame stars distributed among 12 fields for which 
$\Delta\pi_{rel\rightarrow abs}$ was based on both SDSS optical and 2MASS infrared photometry. The
mean difference between these (SDSS -- 2MASS) is +0.10 $\pm$ 0.08 (std dev mean) mas. Thus, there is
no significant systematic difference in the reference frame distances estimated from SDSS and 2MASS.

\section{Astrometric Results}

Table~2 presents our preliminary proper motion and parallax results. The first column gives an abbreviated
object name. The second column gives the spectral type which we adopt (see \S 13). The third column gives
the parallax ($\pi$) solution relative to the reference frame employed along with the standard mean error. The
fourth column gives the absolute parallax, which is corrected for the estimated distances to the reference frame
stars as described in \S 9, along with its standard mean error. The error for the absolute parallaxes
contains, in quadrature, the sum of the error of the relative parallax determination with the error of the absolute parallax
correction. Comparing the relative and absolute parallaxes and their errors demonstrates the small effect of the
distance of the reference frame for these relatively nearby objects. The fifth column gives the relative proper motion 
($\mu$) in mas yr$^{-1}$ with respect to the reference frame along with the formal uncertainty. The sixth column gives
the position angle of the proper motion (in the sense east of north) along with its uncertainty. Using SDSS astrometric
stars in the field for SDSS~J053952$-$0059 for 41 frames spaced over two years we found the mean rotation of the
field as -0.044~$\pm$~0.063 deg. Thus, the natural rotation of ASTROCAM, when pointed at the meridian and
$\delta$~$\approx$~0$^{\circ}$, is indistinguishable from the celestial system, so we have made no rotation correction
to the natural position angle. Finally, in the last column we give the tangential velocity with respect to the
Sun by combining the absolute parallax and relative proper motion results. 

\section{Astrometric Quality}

For the group of 18 objects with an average 21.2 nights and average $\Delta$t = 1.34 yr, the mean parallax error is
4.86 mas and the mean proper motion error is 8.23 mas yr$^{-1}$. For the group of 22 objects with an average 26.7 nights 
and average $\Delta$t = 2.03 yr, the mean parallax error is 3.86 mas and the mean proper motion error is 
5.20 mas yr$^{-1}$. Thus the proper motion errors scale with ($\Delta$t$)^{-1}$ as might be expected if the observations
are uniformly distributed. The parallax errors reduce somewhat faster than if proportional to (nights)$^{-1}$. The most 
likely reason for
this is that the objects with the smaller time baseline do not as yet have observations well distributed over
parallax factor. Certainly the objects with only 1.34 yr of observations must be viewed with some degree of caution
in this regard, although the spectral type versus absolute magnitude figures discussed below would indicate that
there are no gross errors for any object. While we have chosen to not present the Y (declination) parallax results at 
this time, we note that the slope $\pi$(Y)/$\pi$(X) is 0.99~$\pm$~0.05 and that the $\pi$(X) and $\pi$(Y) results for 
all objects agree to within their error bars. 

As Monet et al. (1992) have pointed out, the mean error for a single observation of unit weight (m.e.1) for an ensemble
of stars on frames taken over a period of time is a useful measure of astrometric accuracy. They report a
range of 3 to 5 mas as characteristic of fields observed in the USNO CCD parallax program on the 1.55-m
telescope. For our preliminary data we find mean errors and standard deviations in X and Y, respectively, of
15.5 $\pm$ 4.9 and 17.9 $\pm$ 4.9 mas (8.9 $\pm$ 2.8 and 10.3 $\pm$ 2.9 mas for dithered triplets which are
equivalent to one CCD observation). There is no significant difference between fields observed in J or H. 
These results improve considerably when solutions are run using only the four brightest reference frame stars
with X and Y errors, respectively, of 8.1 $\pm$ 3.4 and 10.0 $\pm$ 4.1 (4.6 $\pm$ 1.9 and 5.8 $\pm$ 2.4 for
dithered triplets). There is no obvious reason why the Y errors are systematically larger than X.

There are several likely reasons why the infrared astrometric accuracies are somewhat 
worse than for CCD results at the same telescope. The first is that we have thus far not solved for proper
motions for any of the reference frame stars. While largely orthogonal to the parallax solution, allowing
for proper motions in the reference frame should significantly reduce m.e.1. Second, we have not made corrections
for DCR, although, since our observations are in the infrared and are constrained to the meridian, we believe this
has little effect. A third issue is the quality of some frames we have had to employ to produce a preliminary
set of results. Figure~1 shows the histogram of seeing at the beginning of each set of observations employed in 
this paper. Although the median image size is 1.33 arcsec (FWHM), we have had to employ many frames with much
worse seeing. Experience from the USNO CCD program has shown that astrometric quality is not significantly degraded 
when exposure times are increased  by (image size)$^2$ so as to maintain a central image density. However, due
to the small effective aperture of our telescope, our exposure times are typically 30 minutes for three dithers
even in the best seeing. Thus, we can afford to only double exposure times in worse seeing. While the poorer
seeing frames, at this point, help the parallax solution, they clearly hurt the m.e.1 statistics. As we accumulate
more data over a longer time baseline the poorer seeing observations will be retired. 

It is instructive to compare our infrared astrometric accuracy with that found by TBK03. Using the European Southern 
Observatory (ESO) 3.5-m New Technology Telescope (NTT) with 0.815 arcsec 
FWHM median seeing, they find 
an m.e.1 of 12.1 mas (0.042 pixel) for each 2 minute individual frame. From above we find an XY--averaged m.e.1 of 
16.7 mas (0.046 pixel) for each typical 10 minute individual frame using a 1.55-m telescope with 1.33 arcsec
FWHM median seeing.

\section{Comparison of $\pi$ and $\mu$ to Previous Results}

\subsection{USNO CCD Program} 

Table~3 compares the results for the seven objects in common between the USNO infrared and optical CCD
programs \citep{dah02}. Column~1 gives the abbreviated object names, the second column whether CCD,
IR, or difference values, and the third column the range of time ($\Delta T$) over which the observations were
observed. The fourth column gives the derived absolute parallaxes, errors, and differences. While it is true that
the CCD and IR parallaxes are derived using different reference frames, correction to absolute parallax
should give consistent results. The fifth and sixth columns give the relative proper motions and
position angles of proper motion and their errors, respectively, along with differences.

There are two objects that have marginally different parallaxes. One is 2MASS~J085035+1057, which is a
binary of 0.16 arcsec separation \citep[discussed in \S 15.4]{rei01} and for which the IR and 
CCD programs also derive marginally different proper motion position angles. Using our IR parallax,
0.16 arcsec semi-major axis, and the maximum mass of the system \citep{rei01} we estimate a minimum orbital
period of the system of $\ge$~51 yr. Given the short $\Delta T$s of both programs it is unlikely that
orbital motion of the system could lead to major errors in the parallax or proper motion determinations.
Also, our analysis in \S 15.4 indicates similar spectral types for the two binary members, so that the photometric 
barycenters should not be strongly affected in either the optical or infrared. More observations in both programs
will be needed of this system to ensure that the derived parallax is without systematic errors.

The second object with marginally different parallaxes is SDSS~J125453$-$0122. There is no indication that
this object is in a binary system. We see in \S 12.2 that the USNO IR parallax is nearly identical
with that measured by TBK03, so we favor the USNO IR parallax over the USNO CCD parallax. (We note that the 
USNO optical measurement is based on only $\Delta T$~=~1.2 yr, so the discrepency may just be due to a small
$\Delta T$.)

The last line of Table~3 gives the weighted mean differences in parallax, proper motion, and proper motion
position angle between the USNO IR and CCD determinations. There is no strong evidence for any systematic
differences in these quantities determined between the two programs.

\subsection{ESO 3.5-m NTT Infrared Program} 

Earlier (\S 11) we discussed the differences in astrometric quality between the USNO infrared
program and the infrared program that has been carried out at the ESO 3.5-m NTT (TBK03). Table~4 compares the parallax 
and proper motion results for the seven objects in common between the infrared programs at USNO and at ESO.
Column~1 gives the abbreviated object names, the second column whether USNO, ESO,  or difference values, and the 
third column the range of time ($\Delta$t) over which the observations were obtained. Since TBK03
did not apply corrections to absolute parallax, in the fourth column we compare relative
parallaxes and their errors. The fifth and sixth colummns give the relative proper motions and position angles 
of proper motion and their errors, respectively.

One of the objects, 2MASS J122554$-$2739AB, is in a binary system (see \S 15.6); however, there is no significant
difference between the parallaxes, proper motions, and position angles determined by both programs. The only
marginally significant differences are for both the parallax and proper motion of 2MASS J104753+2124
and the parallax of 2MASS J121711--0311. It is not
possible to understand which determination is better at this time, although we note that, for this object and
all others in Table~4, the USNO observations are all in the shorter time baseline group of
$\Delta$t~$\approx$~1.3~yr.

The last line of Table~4 gives the weighted mean differences in parallax, proper motion, and proper motion
position angle between the USNO and ESO determinations. There is no evidence for any systematic differences in 
these quantities determined between the two programs.

\section{Adopted Spectral Types and Infrared Photometry}

We compile here the spectral types and infrared photometry, primarily from the literature, to be used in the
ensuing discussion.
In Table 5 we present a small amount of USNO infrared photometry relevant to the objects discussed in this
paper. The data are an addendum to the USNO photometry obtained with IRCAM presented by D02 which gives 
details of the observations and reductions. These data are essentially on the CIT photometric system 
as they were obtained by normalization to the Elias et al. (1982)--based standards of Guetter et al. (2003).
The exceptions to this are the data for SDSS~J020742+0000, for which we used 2MASS All-Sky Point Source Catalog
(PSC) photometry of stars
within the field of view of ASTROCAM, converted those to CIT system values via the transformations of
Carpenter (2001), and used instrumental ASTROCAM $JHK$ magnitudes to obtain $JHK$ photometry for this object.
This process was adopted because 2MASS did not obtain photometry for this faint object and there are no published
empirical transformations between photometry on other systems and the CIT system for T dwarfs. Stephens \&
Leggett (2004) have published synthetic transformations but in this paper we chose to employ empirical transformations
only (see below).

In Table 6 we present adopted spectral types and combined $JHK$ photometry on the CIT system 
\citep{eli82} for the 40 objects presented in this paper. Column 1 gives the abbreviated object name, column 2
the adopted spectral type, and column 3 references for the spectral types, including the discovery reference. 
For L dwarfs we have adopted spectral types based on the optical spectrum classification system of
Kirkpatrick et al. (2000), supplemented by work from several other authors, while for T dwarfs we have adopted
spectral types based 
on the infrared spectrum classification system of Burgasser et al. (2002a), supplemented by the system of 
Geballe et al.~(2002). The three exceptions to this are for the three L dwarfs SDSS~J003259+1410, SDSS~J010752+0041, 
and SDSS~J144600+0024, for which no optical classifications have been published. For SDSS~J083008+4828, we use an
unpublished optical classification from Kirkpatrick et al. (2004). We note that these choices leave a gap between
L8 and T0 as there are, so far, no L dwarfs later than L8 so far classified in the Kirkpatrick et al. (2000) system.
See \S 15 for further discussion of spectral types for several objects of special interest.

The $JHK$ photometry given in columns 4 through 6 is combined from several sources: the 2MASS All-Sky PSC 
({\bf www.ipac.caltech.edu/2mass}) transformed to the CIT system by the relations of Carpenter (2001), for L--dwarfs 
MKO photometry \citep{leg02} transformed to CIT by the relations of Hawarden et al. (2001), previously published 
USNO photometry \citep{dah02} which is on the CIT system \citep{gue03}, UKIRT photometry for four objects \citep{leg00} 
transformed to the CIT system by the relations of Hawarden et al. (2001), and the small amount of new USNO CIT system 
photometry presented in Table 5. The listed photometric values are weighted mean values based on the published
photometric errors after transformation to the CIT system. Column 7 gives references to the photometry employed.

For the purposes of this paper we have chosen to employ published empirical transformations from the various systems
to the CIT system. Stephens and Leggett (2004) have pointed out the potential dangers in using transformations based
on normal stars for L and T dwarfs and have calculated synthetic transformations from various systems to the MKO
system. Examination of their results show that the predicted systematic errors are only a few percent for all
transformations, except for the 2MASS to CIT transformation at K band which are approximately 0.05 mag for L dwarfs
and ranging from 0.10 mag for early and mid T dwarfs to as large as 0.20 mag for the latest T dwarfs.
However, 2MASS photometry for the late T dwarfs studied here is either unavailable or with intrinsic
photometric errors at the 0.2 mag level. Thus, random errors, along with the reported 0.05 to 0.25 mag intrinsic variations 
for these objects (Enoch, Brown, \& Burgasser 2003) dominate this potential source of systematic error.

\section{Discussion}

In this section we present selected infrared absolute magnitude versus spectral type and infrared color
relationships. At the risk of being accused of astronomical chauvinism, we have chosen for the remainder of
this paper to discuss only USNO--derived optical and infrared parallaxes (this paper and D02). This 
provides a self--consistent set of parallax and proper motion determinations using the same telescope and similar
observing philosophies and reduction software, with only the detector being different. 
When we publish completed parallax solutions, it will then be appropriate to combine these results
with those of other researchers. Four of the objects, 2MASS~J085035+1057AB, 2MASS~J122554$-$2739AB, 2MASS~J172811+3948AB,
and 2MASS~J210115+1736AB  are known binaries. We discuss how the separated spectral types and photometry are determined 
in \S 15.4, \S 15.6, \S 15.8, and \S 15.9, respectively.

\subsection{Absolute Magnitude versus Spectral Type}

In Figure~2 we plot J-band absolute magnitude ($M_J$) versus spectral type. The solid data points are the results from this
paper, where we have combined the infrared parallaxes with the infrared photometry and spectral classifications listed
in Table~6. The photometric errors have been convolved with the parallax uncertainties to produce the vertical
error bars. The horizontal errors are $\pm$ 0.5 spectral type for those objects with well-determined spectral
classification and $\pm$ 1.0 spectral type for those with less certain classifications. The open data points are
from D02 where we have used the parallaxes, infrared photometry, and spectral types published
in that paper. For the seven objects in common between D02 and this paper, we plot both the CCD- and
infrared-derived absolute magnitudes using the photometry of Table~6. In order to be consistent photometrically,
we plot T513$-$46546 (D02) using 2MASS All-Sky PSC photometry transformed to CIT values by the Carpenter (2001) transformations.
Several of the objects, in both the infrared and optical parallax groups, are known binaries and we have plotted
them in accordance with what is known about their binary natures. These objects are discussed individually in
\S~15 and \S~16 below. In Figures 3 and 4 we plot the $M_H$ and $M_K$ absolute magnitudes, respectively, versus spectral 
type with the same considerations as for Figure~2.  

Our results for spectral types earlier than about L5 do not provide much new information, since we have only a few
early L dwarfs, some of which currently have large error bars. However, the results are consistent with narrow loci in all
three diagrams. For objects between L5 and L9 the dispersion is clearly much greater than for earlier objects. The 
widths are about 1.5 mag, 1.3 mag, and 1.0 mag in J, H, and K, respectively, for the L5--8 objects.  While this could be due to
an admixture of objects of different ages, masses, and gravities, we have looked, within a given spectral type,
for correlations of $M_J$, $M_H$, and $M_K$ with tangential velocities (see \S 14.3) as a potential age indicator, but 
have found none. More likely, the additional width is due to the complicated atmospheric physics expected for late L dwarfs
\citep{bur02b,ste03}, which also can explain why the spread is a function of wavelength. These models also predict significant 
variability due to rapid evolution or motion of cloud holes, which alone could be responsible for the apparent spread in
absolute magnitude. Clearly photometric monitoring will be necessary to fully understand the L--T transition objects.

In all three diagrams the transition from L to T dwarfs is smooth. The excess in luminosity for T1-5 spectral types, 
previously noted by D02 and TBK03, is clearly substantiated by our enhanced database. 
While this excess is most evident in the J band, it is also seen in the H and K bands.
The possibility that the early T dwarf luminosity excess is caused by contamination due to binary systems \citep{bur01} 
now seems unlikely due to the sheer number of objects which participate in this hump. TBK03 point out that the amplitude of 
the hump, like the spread at late L, is also unlikely to be explained by an age selection effect \citep{tsu03}. 
T dwarfs between T6 and T8 once again form a rather tight locus terminating at $M_{J,H,K}$~$\approx$~16.6. 

The additional data for T dwarfs allow a somewhat clearer picture of the L--T transition region in these
diagrams. While there is no self--evident reason to believe that $M_J$, $M_H$, or $M_K$ should map linearly
with spectral type (TBK03), we note that inspection of Figures 2 through 4 shows that the late T dwarfs (T6--T8) are on 
a rough extension of the absolute magnitude versus spectral type relation of the the early L dwarfs (L0--L5). Relative
to a fiducial line drawn between the early Ls and late Ts, in J--band the L--T transition objects show a 
luminosity deficit of about 1.5 mag at L6--L8 and a luminosity excess of about 1.5 mag at T1--T5. In H--band
the L6--L8 deficit has shrunk to about 0.5 mag and disappears at K--band, while in H-- and K--bands the excess
for T1--T5 objects remains at 1.0--1.5 mag. 

We note that the contrast between the large spread of absolute magnitudes at late L versus the apparently narrow
locus of early T absolute magnitudes may not be significant. The narrow T dwarf locus may yet prove to be an
artifact of small number statistics. Also, spectral typing may just map out an equivalent diversity of physics over
a smaller range of spectral types at late L than at early T.

\subsection{Absolute Magnitude versus Infrared Colors}

In Figure~5 we plot $M_J$ versus $J-H$ color. The solid data points are the results from this
paper, where we have combined the infrared parallaxes with the infrared photometry and spectral classifications listed
in Table~6. The photometric errors have been convolved with the parallax uncertainties to produce the vertical
error bars, while the horizontal errors are from Table~6. The open data points are those from D02, where we 
have used the parallaxes, infrared photometry, and spectral types published in that paper. We treat the seven objects 
in common between D02 and this paper and the binary systems as described in \S 14.1.
In Figure~6 we plot $M_J$ versus $J-K$ color, while in Figures 7 and 8 we plot $M_K$ versus $J-H$ and $J-K$,
respectively.

These figures show the well-known color trends for L dwarfs ranging from $J-H$~$\approx$~0.7, $J-K$~$\approx$~1.1 for
L0 ($M_J$~$\approx$11.0, $M_K$~$\approx$~10.0) to $J-H$~$\approx$~1.2, $J-K$~$\approx$~2.1 for L8 ($M_J$~$\approx$~15.0,
$M_K$~$\approx$~13.5). Late T dwarfs (T5 to T8) all have roughly $J-H$~$\approx$~0, $J-K$~$\approx$~0, 
while early T dwarfs (T0.5 to T4.5) have a wide range of colors transitioning between late L and late T colors.
Several objects are now placed in the transition region between the loci of the L and T dwarfs, which is best shown in 
the $M_K$ versus $J-K$ diagram (Figure~8). Unfortunately, the locations of many of the transition objects and early T dwarfs 
are poorly known at this time due to uncertain distances. We will defer comparing evolutionary models with observations until 
we obtain final parallaxes and further USNO-CIT photometry. However, we note that the apparent brightening in 
$M_J$ across the L--T transition is consistent with the predictions of the cloud hole model of Burgasser et al. (2002b),
while the fact that the brightening is an apparent trend argues against the hypothesis of Tsuji \& Nakajima (2003)
that it is simply an age effect.

\subsection{Kinematics}

For stars in the solar vicinity, motion with respect to the Sun is an indicator of age, since older stars will 
have had time to be perturbed preferentially to different orbits by interaction with the Galactic disk. Because
T dwarfs are thought to be the cooler and older analogs of at least some L dwarfs, it might be expected that
T dwarfs will have a larger mean velocity, with respect to the Sun, than L dwarfs. The measured tangential velocities 
(V$_{tan}$) with respect to the Sun measured primarily for L dwarfs in D02 can be combined with those 
for L and T dwarfs in this paper (Table~2) to have sufficient objects in order to compare the velocity distributions.
For the seven objects in common between the two papers we use the weighted mean values of V$_{tan}$. We remove
three objects from consideration with exceptionally large V$_{tan}$ uncertainties, having both 
$\sigma$(V$_{tan}$)~$>$~10~km~s$^{-1}$ and V$_{tan}$/$\sigma$(V$_{tan}$)~$<$~3 (two L dwarfs: 
2MASS J143535$-$0043 [31.7$\pm$13.3 km s$^{-1}$] and 2MASS J095105+3558 [55.8$\pm$32.7 km s$^{-1}$]
and one T dwarf: SDSS~J083717$-$0000 [24.3$\pm$11.8 km s$^{-1}$]; all from this paper). This leaves 33 L 
dwarfs and 17 T dwarfs to make the velocity comparison. The unweighted average values of tangential velocity for L and T
dwarfs, respectively, is 30.0~$\pm$~3.6 and 43.0~$\pm$~4.8 km s$^{-1}$. The median values are, for L and T dwarfs
respectively, 24.5 and 39.0 km s$^{-1}$. As D02 have discussed, the velocity of the L dwarfs is
consistent with velocities for old disk M and dM stars.

The distributions of V$_{tan}$ are shown in Figure~9 for L dwarfs in the top panel and for T dwarfs in the bottom panel. 
Based on the Kolmogorov--Smirnov test, the null hypothesis that the two distributions are indistinguishable can only
be rejected at the 73\% level. The main difference is that the T dwarfs have no examples with
V$_{tan}$~$\le$~20 km s$^{-1}$, whereas the L dwarfs have 11 of 33 (33\%) with V$_{tan}$~$\le$~20 km s$^{-1}$. While
there are fewer objects in the T dwarf subset, if the distributions were the same, 5.7 T dwarfs with
V$_{tan}$~$\le$~20 km s$^{-1}$ would be expected. 

\section{Discussion of Individual Objects}

There are several objects which, after our new parallax results, warrant special attention. We discuss these
briefly here.

\subsection{2MASS J041519$-$0935}

This object is the assigned T8 standard on the Burgasser et al. (2002a) classification system, which we have adopted 
as the primary classification system for T dwarfs in this paper. However, in the classification system of Geballe (2002) 
it is T9 \citep{kna04}, indicating that it is likely the latest spectral type T dwarf yet found. Although it makes
little quantitative difference, we plot and fit the the object as a T8.5, to recognize its extremely late spectral
type. With a parallax of 174.3~$\pm$2.8 mas (5.74~$\pm$0.09~pc) it is one of the closest known brown dwarfs.  
Combining this distance with its 2MASS photometry gives it the lowest
absolute magnitude of any T dwarf known: $M_J$~=~16.92~$\pm$0.07 (see \S 17). We estimate a bolometric magnitude of
$M_{bol}$~=~18.70~$\pm$~.26, a luminosity of log(L/L$_{\odot}$)~=~--5.58~$\pm$~0.10, and, hence,
a temperature of T$_{\rm eff}$~=~760~$\pm$~80 K, making it the least luminous and coldest brown dwarf yet discovered. 
See \S 18 for our discussion of bolometric corrections and temperature estimates for T dwarfs.

\subsection{SDSS J042348$-$0414}

The L--T transition object SDSS~J042348$-$0414 has unusual spectral properties such that in the near--infrared
Geballe et al. (2002) classify it as a T0, while in the optical Cruz et al. (2003) and Kirkpatrick et al. (2004) classify
it as an L7.5. Both classifications are correct in the wavelength regions considered. It has a K--band bolometric 
correction which is typical of L7.5 to T0 objects \citep{gol04}.  At L7.5 it lies at the top
of the loci for late L dwarfs in the spectral type versus absolute magnitude diagrams of Figures 2
through 4, about a factor of two brighter than the average for late L dwarfs. At T0 it lies well above the locus for
early T dwarfs in these same diagrams; about a factor of 4 brighter than the average for early T dwarfs. While
this could be explained by binarity (multiplicity in the case of T0), no HST observations have yet been obtained to
help resolve this issue. While we have not yet carried out a formal search for perturbations,
there is no evidence in the astrometric data to indicate either perturbations or difficulty
in centroiding for this nearby (15.2 pc) object. From \S 11, for the group of 22 objects with average $\Delta$t~=~2.03 yr, 
the mean parallax error is 3.86 mas and the mean proper motion error is 5.20 mas yr$^{-1}$. SDSS~J042348$-$0414 has 
$\Delta$t~=~2.02 yr, with mean parallax and proper motion errors of 1.70 mas and 2.80 mas yr$^{-1}$, respectively. 
Another possibility is that this object is still very young and thus over--luminous due to a radius larger than
for older brown dwarfs. For the purposes of this paper we will use the L7.5 classification since it lies within the 
absolute magnitude loci of late L dwarfs and its near--infrared colors are not inconsistent with this spectral classification. 
Clearly further investigation will be needed to understand the nature of this enigmatic
object.

\subsection{2MASS J055919$-$1404}

2MASS J055919$-$1404 has historically been considered enigmatic since it appeared to have an absolute
magnitude considerably larger than objects of similar spectral type. Although it has been suspected
of being an equal mass binary \citep{bur01,dah02}, recent Hubble Space Telescope Wide Field/Planetary
Camera 2 (WFPC2) observations have failed to reveal a bright binary companion at a separation larger than 
0.05 arcsec \citep{bur03a}, although such a system could currently be hidden in an unfavorable orientation.
Our data adds considerably to the census of T1--5 ojects which populate the hump seen in the spectral type
versus absolute magnitude diagrams (Figures 2--4). These results indicate that 2MASS J055919$-$1404 lies 
at least much closer to the loci of other objects in this spectral range than was previously thought. It 
will require further luminosity determinations of other T1--5 objects and other observations
of 2MASS J055919$-$1404 itself to determine whether this object is indeed overluminous or simply the prototype hump
object lying at the peak of this local luminosity maximum.

\subsection{2MASS J085035+1057AB}

2MASS J085035+1057AB was found by Reid et al. (2001) to be a binary system with a separation
of 0.16 arcsec from HST WFPC2 observations. Based on optical colors from 
these observations they estimate a J--band magnitude difference of $\Delta$M~=~0.9. Adopting this 
magnitude difference, the photometry listed in Table~6, and our parallactic distance we find a
value of $M_J$~=~13.78 for the A component, which is consistent with the L6V spectral type
for the dominant component found by Kirkpatrick et al. (1999, see Figure~2). The B component
has $M_J$~=~14.68, which is consistent with a late-L, early-T, or a T6 spectral type (again see
Figure~2). A mid to late T spectral type can probably be ruled out on the basis of the
combined spectrum \citep{rei01}. Since late--L dwarf colors range between $(J-K)$~$\approx$~1.6--2.0 and early-T colors 
cluster around $(J-K)$~$\approx$~1.0, measurement of a K magnitude alone would likely distinguish between
a late-L and early-T spectral type for the B component.

\subsection{2MASS J093734+2931}

The second faintest T dwarf in our sample, and one of the closest (6.14 $\pm$ 0.15 pc), is 2MASS~093734+2931,
which was classified as a peculiar T6 by Burgasser et al. (2002) due to its extremely blue
near--infrared color ($(J-K)$~=~$-$0.72 $\pm$ 0.20; Table~6). This object is indeed peculiar, in that
it is over 3 times less luminous and about 300 K cooler than the typical T6 dwarf observed (see \S 18 and \S 19).
Burgasser et al. (2002, 2003b) postulate that 2MASS~093734+2931 may be a metal--poor and/or high gravity
(i.e. old and massive) brown dwarf based on the enhanced collisionally--induced H$_2$ absorption that
gives it its blue near--infrared colors. These gravity/metallicity effects may be substantial enough
to modify the near--infrared spectrum so that it appears to be an earlier--type (hotter) brown dwarf than
its true T$_{\rm eff}$. Alternatively, K--band bolometric corrections for this object may also be biased,
due to the substantial suppression of flux at these wavelengths (Golimowski et al. 2004). However, given
that this object is subluminous in all near--infrared bands as compared to other T6 dwarfs, metallicity and/or
gravity differences are clearly important in these cool brown dwarfs and must be quantified before a universal spectral
type/temperature relation can be derived.

\subsection{2MASS J122554$-$2739AB}

2MASS J122554$-$2739 is a clearly separated double system with an angular separation of 
0.282~$\pm$~0.005 arcsec from HST WFPC2 observations \citep{bur03a}. 
Based on optical colors from these observations they speculate that the system is
composed of T6V and T8V components and estimate a J--band magnitude difference of 
$\Delta$$M_J$~=~1.35~$\pm$~0.08. Adopting this magnitude difference and assuming that it
also applies to H-- and K--band (since late T dwarf $JHK$ colors scatter around 0.0), we
use the available 2MASS photometry transformed to the CIT system, and apply our
parallactic distance determination to produce the $JHK$ absolute magnitudes listed
in Table 7 (see \S 17) and plotted in Figures 2 through 8. The infrared absolute magnitudes are consistent
with T6V and T8V for components A and B, respectively.

Our parallax result increases the projected separation of the components to 3.80~$\pm$~0.18 AU,
from the 3.18 AU separation based on the spectrophotometric distance \citep{bur03a}.
This, in turn, increases the estimated orbital period of the system to between 31 and 53 years,
based on the effective temperatures and age range assumed by Burgasser et al. (2003a), but
does not alter their conclusion that direct orbital motion of this system could be detected
over a short time range.

\subsection{SDSS J143517$-$0046 and SDSS J143535$-$0043}

SDSS J143517$-$0046 and SDSS J143535$-$0043 were put on the infrared parallax program partially
because, at an angular separation of about 5 arcmin, they could be a common proper motion pair. 
While the current parallax results do not rule out the possibility that they are at the same
distance, the proper motions are nearly a factor of four different in magnitude and the position angles
nearly orthogonal. Thus, the angular proximity of these objects appears to be fortuitous.

\subsection{2MASS~J172811+3948AB}

Gizis et al. (2003) found this to be a binary with 0.13 arcsec separation using the HST WFPC2 camera.
They found the A component to be 0.3 mag brighter than the B component in I band, but 0.3 mag fainter 
in Z band, which they interpret as the B component being an L/T transition object. Since the $J-H$ and $J-K$
colors (Table~6) are normal for the L7 spectral classification \citep{kir00} and absolute $JHK$ magnitudes are
approximately equal for late L to early T, we assume that the flux is evenly split between the two components
in the infrared to produce the $JHK$ absolute magnitudes listed in Table 7 (see \S 17) and plotted in Figures 2 through 8. 

\subsection{2MASS~J210115+1756AB}

Gizis et al. (2003) found this to be a binary with 0.232 arcsec separation using the HST WFPC2 camera.
They find the B component to be slightly later than the L7.5 classification of the system \citep{kir00}.
Since the colors and absolute $JHK$ magnitudes are similar for late L dwarfs, we assume that the flux is evenly split 
between the two components in the infrared to produce the $JHK$ absolute magnitudes listed in Table 7 (see \S 17) and 
plotted in Figures 2 through 8. 

\section{Binaries in the CCD Dataset}

There are four additional binary systems in the CCD astrometry set which we treat in a manner
similar to that of D02. Two systems have components of equal near--infrared magnitudes:
DENIS 020529$-$1159AB \citep{koe99,leg01} and DENIS 122815$-$1547AB \citep{koe99}. For these
systems we assume that the components have identical spectral types and plot the absolute magnitudes
with the assumption that the flux is split evenly between the components.

In the case of 2MASS J074642+2000AB, Reid et al. (2001) estimate a J--band magnitude difference of 
$\Delta$$M_J$~=~0.47 based on optical colors from HST WFPC2 observations. We
use this magnitude difference to estimate the J magnitude for the L0.5 primary component and assume
that the combined light colors apply to the primary to estimate its H and K magnitudes. While there is not enough 
information available to uniquely determine the nature of the B component, we note that its resultant 
$M_J$ = 12.31 $\pm$ 0.02 is consistent with its being about an L3V. Since $JHK$ colors of L0.5 and L3 objects are similar
(Table~9), our assumption that the system $JHK$ colors are applicable to the primary for determining $M_H$ and
M$_K$ is reasonable.

For 2MASS J114634+2230AB, Reid et al. (2001) estimate a J--band magnitude difference of $\Delta$M~=~0.23, 
again based on optical colors from HST WFPC2 observations and we again 
use this magnitude difference to correct the J magnitude for the L3 primary component and assume
that the combined light $JHK$ colors apply to the primary. The small difference in $\Delta$M indicates that
the B component must be near an L3 also and that our assumption that the system $JHK$ colors are applicable 
to the primary for determining $M_H$ and $M_K$ is reasonable.

\section{Absolute $JHK$ Magnitudes Based on Optical and Infrared Parallaxes}

In this section we combine the adopted spectral types and CIT system photometry from Table~6 with
the infrared parallaxes presented in this paper (Table~2) and the optical parallaxes derived by
D02 to derive CIT system absolute magnitudes $M_J$, $M_H$, and $M_K$. These
are presented in Table~7 in order of spectral type. We note that the spectral types used for the objects in D02
conform to the convention for L and T spectral types described in \S 13. In the seven cases where both USNO optical and infrared
parallaxes are available, we combined the parallaxes, weighting by the listed uncertainties. 
The first three columns give the abbreviated object names, adopted spectral type, and whether optical, infrared,
or combined parallaxes were used, respectively. The last six columns give $M_J$, $M_H$, and $M_K$ and their
uncertainties, respectively. We have not included the B components of binaries where spectral types are uncertain.

For the 36 objects from L0 to L8 (excluding the enigmatic object SDSS~J042348$-$0414) we use the data of Table~7 to 
derive the following equations giving the best linear fits of $M_J$, $M_H$, and $M_K$, respectively,
with spectral type (ST), where we numerically encode ST$_L$~=~0 for L0, ST$_L$~=~5 for L5, etc. There is no 
evidence for significant
second order terms for any of the bandpasses. We weighted the data by the inverse square of the larger of the 
$\sigma$($-$)$\sigma$(+) values and, since these are all `preliminary' parallaxes degraded the minimum error to
0.05 mag such that the few objects with very small formal errors would not dominate the fits. The uncertainties
listed at the end of each equation show the dispersions from the fits, in magnitudes.
\begin{equation}
M_J = 11.62 + 0.380 \times (ST{_L}), \sigma = 0.24
\end{equation}
\begin{equation}
M_H = 10.85 + 0.346 \times (ST{_L}), \sigma = 0.19
\end{equation}
\begin{equation}
M_K = 10.33 + 0.324 \times (ST{_L}), \sigma = 0.17
\end{equation}
The changing slopes with bandpass reflect the fact that the L dwarfs become redder with later spectral type.
Our results are consistent with those of D02 who report an $M_J$ slope of 0.347 and scatter of $\sigma$~=~0.24 mag 
for objects between M6.5V and L8. 

For the 19 T dwarfs between T0.5 and T8 the following quadratic equations adequately describe the $M_J$, $M_H$, and $M_K$ 
versus spectral relations where ST$_T$~=~0 for T0, ST$_T$~=~5 for T5, etc. In this case we have degraded the minimum
uncertainty to 0.10 mag so that essentially one object would not dominate the fit.
\begin{equation}
M_J = 15.04 - 0.533 \times (ST{_T}) + 0.091 \times (ST{_T})^2, \sigma = 0.36
\end{equation}
\begin{equation}
M_H = 13.66 - 0.139 \times (ST{_T}) + 0.063 \times (ST{_T})^2, \sigma = 0.44
\end{equation}
\begin{equation}
M_K = 13.22 - 0.055 \times (ST{_T}) + 0.060 \times (ST{_T})^2, \sigma = 0.62
\end{equation}

\section{L and T Dwarf Luminosities and Effective Temperature Estimates}

L and T dwarf luminosities and effective temperatures can be estimated for objects with known distances.
D02 have previously made such estimates for mid-M stars through late L dwarfs, but this has been difficult
to do for T dwarfs due to the paucity of bolometric corrections (BC$_{filt}$~=~$M_{bol}-M_{filt}$)
which have previously been measured for these objects (e.g. for 2MASS J055919$-$1404
[\citealp{bur01}]; for Gliese 229B [\citealp{leg99,leg02}]; for Gliese 570D [\citealp{bur00,geb01,leg02}]). 
However,  Golimowski et al. (2004) have recently calculated BC$_K$ for a large number of both L and T dwarfs, 
greatly expanding the number of consistently--derived bolometric corrections available for these objects.  
Moreover, they have included L$^{\prime}$ and M$^{\prime}$ photometry in their BC calculations, such that most of 
the spectral energy distributions of L and T dwarfs are now
covered for the first time, allowing more reliable estimates of bolometric corrections and, hence,
bolometric magnitudes to be made. 

We use the $M_K$ values from Table~7 and apply the fourth order polynomial fit of BC$_K$ versus spectral type from 
Table~4 of Golimowski et al. (2004) to determine $M_{bol}$ for L and T dwarfs which have USNO parallaxes.
The resultant $M_{bol}$ values, along with their balanced uncertainties, are listed in column 3 of Table~8. We 
note that the $M_K$ values are on the CIT system, while the Golimowski et al. (2004) BC$_K$ determinations are formally
for the $K_{MKO}$ filter. However, the synthetic transformations of Stephens and Leggett (2004) show that the CIT and
MKO filters give nearly identical results for L and T dwarfs, with offsets of 0.01-0.02 mag for L0 to L8 and 0.02-0.09
mag for T0.5 to T8. These small offsets are dominated by other sources of uncertainty but are added to the error budget. 
Also the Golimowski et al. (2004) BC$_K$ versus spectral type calibration is for the spectral classification system of Geballe
et al. (2002), which includes L and T spectral types as late as L9 and T9, but we expect no significant effects from
this. The errors listed for $M_{bol}$ are the quadrature sums of the following uncertanties: 0.02$\le$$\sigma$($M_K$)$\le$1.28
mag, 0.01$\le$$\sigma$($K_{\rm CIT}-K_{\rm MKO}$)$\le$0.09 mag, and $\sigma$(BC)~=~0.13 mag.  
Assuming for the Sun $M_{bol}$~=~+4.74 \citep{dri00}, we list the derived logarithmic luminosities
in units of solar luminosity along with their uncertainties in column 4 of Table~8.

Effective temperatures (T$_{\rm eff}$) for L and T dwarfs can be estimated from L~=~4$\pi$R$^2$$\sigma$T$_{\rm eff}^4$, 
normalized to solar units, by
\begin{equation}
M_{bol} = 42.36 - 5 \log (R/R_{\odot}) - 10 \log T_{\rm eff} 
\end{equation}
\citep{dri00}. L and T dwarf radii are largely, but not totally, independent of mass and age, with a range of about
30\%. Burgasser (2001) performed a Monte Carlo analysis of the Burrows et al. (1997) L and T dwarf
evolutionary models, to understand the distribution of radii for objects between 0.001 and 10 Gyr in age and masses
between 1.0 and 100 M$_{\rm JUP}$. The mean radius for the model assuming constant birth rate and the mass function
$dN \propto M^{-1}dM$ is 0.90 $R_{\rm JUP}$. A total radius range of about 0.75 to 1.05 $R_{\rm JUP}$ results from 
these simulations. We use a radius of 0.90~$\pm$~0.15 $R_{\rm JUP}$ and the previously derived values of $M_{bol}$ and 
its uncertainties to derive the T$_{\rm eff}$ values and their range which are listed in the last column of Table~8.
At log(L/L$_{\odot}$)~=~-5.58 $\pm$ 0.10 and T$_{\rm eff}$~$\approx$~760 K, 2MASS J041519$-$0935 is found to be the
coldest and least luminous brown dwarf yet discovered.

While luminosities and temperatures drop quickly with spectral type for early L and late T objects,
the derived values of $M_{\rm bol}$ and log(L/L$_{\odot}$) show only a small range between about L6 to T5.
When combined with the assumptions about brown dwarf radii above, this implies a temperature range of only about 
1200 to 1550 K. A similar narrow range in T$_{\rm eff}$ was predicted by Kirkpatrick et al. (2000) based on
luminosity estimates for Gliese~584C (L8) and Gliese~229B (T6.5). Such a narrow T$_{\rm eff}$ range over a
substantial shift in spectral morphology is predicted by Burgasser et al. (2000b), who postulate that the L--T
transition is dominated by condensate cloud evolution, rather than cooling. However, more recent cloud models
(e.g. Tsuji \& Nakajima 2003) also predict a rapid evolution from L to T without the need for atmospheric
dynamics.  Clearly, synoptic monitoring, over a range of timescales, both photometrically and spectroscopically, will 
need to be a high priority in order to understand better the competition between cloud condensation and turbulent 
cloud disruption in these objects. 

We note that our derived temperatures for the earliest L dwarfs of about 2400-2500 K are warmer by about
200-300 K than some earlier estimates (e.g. Leggett et al. 2001,2002) but consistent with those derived
by D02. We again direct attention to Golimowski et al. (2004), not only for their definitive BC calculations, but 
also for determinations of luminosities and temperatures for a somewhat
different sample of objects, using a different photometric database and slightly different assumptions about brown dwarf radii.

\section{Mean Derived Properties of L and T Dwarfs}

In Table~9 we present mean derived properties of brown dwarfs based on the work of this paper.
The spectral type ranges listed in column~1 (L0 to L8 and T0.5 to T8) span those of objects with USNO optical or infrared
parallaxes. $M_J$, $M_H$, and $M_K$ listed in columns 2, 3, 4, respectively, are calculated from equations (1) through (6)
of \S 17, while the $J-H$ and $J-K$ colors in columns 5 and 6, respectively, are differences from these equations.
$M_{\rm bol}$, log(L/L$_{\odot}$), and T$_{\rm eff}$ in columns 7, 8, and 9, respectively, are derived from the listed
$M_K$ values and the methods discussed in \S 18. T$_{\rm eff}$ values are rounded to the nearest 10~K. We caution
that the values in Table~9 should be treated as schematic results only, as they do not represent the full width and
subtle variations in the spectral type versus absolute magnitude relations of Figures 2 through 4. With better
astrometry, photometry, and knowledge of brown dwarf variability some of the apparent details in these figures may prove
to be false, while others may prove to represent stellar and sub--stellar physics.

\section{Future Work}

As with the USNO optical parallax and proper motion program, it is our intent that the infrared
astrometry program will be a long--term enterprise. All of the objects listed in this paper will be observed
for a minimum of three years, to ensure a full distribution of observations over the parallactic ellipse, before
final astrometric results are reported. Fainter objects and/or those with poorer reference frames will need a
longer period of time. Our goal will be to provide 1.0 mas or better parallaxes for as many objects as possible,
although it may not be a feasible goal for all objects. Objects with special astrophysical significance may be
left on the program longer to achieve even higher precision. 

It can be anticipated that many more brown dwarfs will be found by further mining of large--sky databases. As objects are 
removed from the current program as they are completed, new objects will be added to the program when they can fit into a depleted
right ascension slot. We have already expanded our program from the initial 40 objects to 52 objects. Most of the new
objects are T dwarfs as we wish to utilize our resources on objects that are observed to advantage in the infrared
while problematic for the optical CCD program. From experience with our current observing mode, a program
consisting of 50 to 60  objects is a practical limit. We intend to publish an update to our results about every 
two years, providing lists both of completed astrometry and preliminary results for new objects and objects with continuing
observations.  

A comparison of uncertainties listed in Table~6 and Table~7 shows that for several objects the uncertainty in absolute
magnitude is largely due to uncertainty in infrared photometry. We intend to provide high quality USNO-CIT photometry,
as in D02, for all objects for which we publish completed astrometry. These frames, along with the
potential to obtain SDSS optical photometry from Flagstaff, will assure that photometric parallaxes of all suitable
reference frame stars will be available to form the best possible relative to absolute parallax corrections.

An interesting related issue is the question of how much of the spread in brown dwarf 
loci in the spectral type and color versus magnitude diagrams is due, not to mean intrinsic brightness dispersion, 
but to variability, especially for the L--T transition objects. The results of a several night 
photometric monitoring campaign during one month (Enoch, Brown, \& Burgasser 2003) show variability for 9 L and T dwarfs 
of 5--25\% in K$_s$ band. Our astrometric database should prove to be 
valuable for investigation of J band variability of T dwarfs and H band variablity of L dwarfs on time scales of
minutes (one night's data), days (one run's data), months (one year's data), and years (an entire astrometry dataset).

\section{Summary}

In this paper we have presented preliminary parallaxes and proper motions for 22 L dwarfs and 18 T dwarfs
derived over time baselines of only $\Delta$t~$\approx$~1.3 or $\Delta$t~$\approx$~2.0 years. The resultant
mean parallax uncertainties of 4.86 and 3.85 mas, respectively, which will be greatly improved by ongoing further
observations, are nonetheless of sufficient quality to provide some significant new results for T dwarfs. We list
here a summary of the more important conclusions we reach from our work.

A. The luminosity excess `hump' for early to mid T dwarfs in the absolute magnitude versus spectral type diagram is 
clearly confirmed. While seen most strongly at J--band, it is also evident in the H-- and K--bands.
The possibility that the hump is due to a selection effect of binaries is likely ruled out by the 
large number of objects participating in the hump.  

B. L5--L8 dwarfs have a significantly larger spread in the absolute magnitude versus spectral 
type diagrams than do earlier L dwarfs. 

C. Late T dwarfs have a narrow locus in absolute magnitude versus spectral type diagrams, similar to the early
L dwarfs. Relative to a straight line connecting the earlier and later objects, the L--T transition objects show systematic
trends.  In J--band the late L dwarfs show a luminosity deficit and the early T dwarfs a luminosity excess.  
The late L dwarf luminosity deficit is less in H--band and is gone in K band, while the early T dwarf luminosity excess 
amplitude is somewhat less in H--band and K--band than at J--band.

D. The absolute magnitude behavior across the L--T transition described above exemplifies the critical role of condensate
cloud evolution at these temperatures based on the most recent spectral models. 

E. Using newly derived bolometric corrections for L and T dwarfs by Golimowski et al. (2004) we derive luminosities and
T$_{\rm eff}$ for L and T dwarfs with USNO--derived parallaxes either from this paper or D02.

F. 2MASS J041519$-$0935 is found to be the least luminous [log(L/L$_{\odot}$)~=~$-$5.58]
and therefore coldest (T$_{\rm eff}$~$\approx$~760 K ) brown dwarf yet found. 

G. We find a broader distribution of L dwarf tangential velocities compared with that of the T dwarfs. While essentially the
same between 20 to 60 km s$^{-1}$, the T  dwarfs do not have a low velocity population as do the L dwarfs. This is 
consistent with T dwarfs being, in general, older than L dwarfs.

\acknowledgments

This research has made use of the NASA/ IPAC Infrared Science Archive, which
is operated by the Jet Propulsion Laboratory, California Institute of
Technology, under contract with the National Aeronautics and Space
Administration. 

    Funding for the Sloan Digital Sky Survey (SDSS) has been provided by the Alfred P. Sloan Foundation, the Participating
Institutions, the National Aeronautics and Space Administration, the National Science Foundation, the U.S. Department of
Energy, the Japanese Monbukagakusho, and the Max Planck Society. The SDSS Web site is {\bf http://www.sdss.org/}.

    The SDSS is managed by the Astrophysical Research Consortium (ARC) for the Participating Institutions. The
Participating Institutions are The University of Chicago, Fermilab, the Institute for Advanced Study, the Japan
Participation Group, The Johns Hopkins University, Los Alamos National Laboratory, the Max-Planck-Institute for Astronomy
(MPIA), the Max-Planck-Institute for Astrophysics (MPA), New Mexico State University, University of Pittsburgh, Princeton
University, the United States Naval Observatory, and the University of Washington.

We thank C. Dahn, H. Harris, and D. Monet for many helpful science discussions.
We thank A. Hoffmann and the team at RVS for working with us to build ALADDIN arrays, the design of which was partially
funded for this work. We thank D. Toomey and his team at Mauna Kea Infared and J. Fischer and the team at the Naval 
Research Laboratory for designing and fabricating our astrometric imager ASTROCAM. AJB acknowledges support provided
by NASA through Hubble Fellowship grant HST-HF-01137.01 awarded by the Space Telescope Science Institute, which is
operated by the Association of Universities for Research in Astronomy, Inc., under NASA contract NAS 5-26555.

\clearpage

\begin{figure}
\plotone{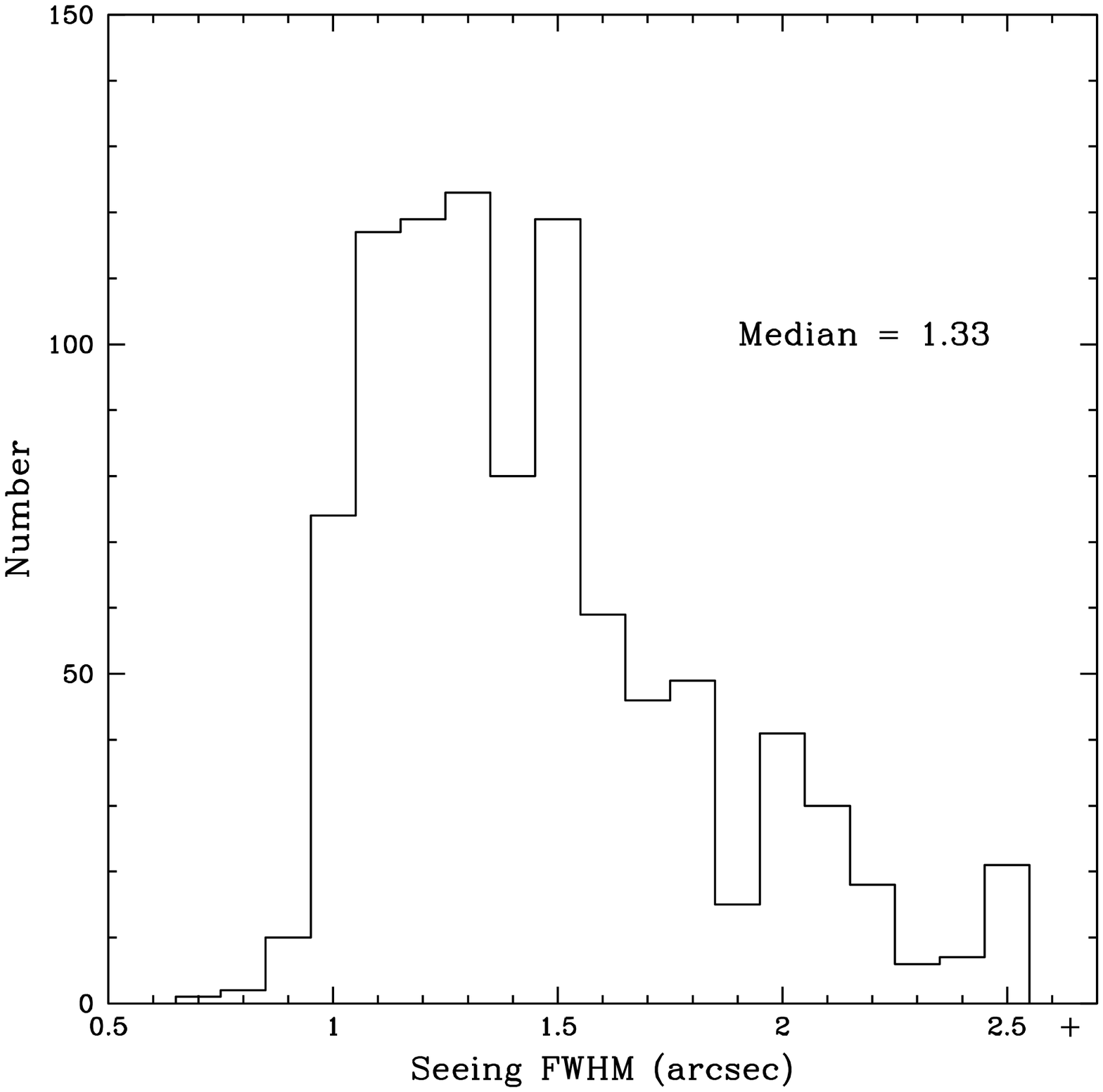}
\caption{Histogram of the seeing (in the J and H bands) at the 1.55-m telescope using ASTROCAM at the beginning of each set 
of observations employed in this paper. The bar at 2.5+ arcsec represents those observations at 2.5 arcsec
and slightly larger used in the solutions. The median seeing is 1.33 arcsec. \label{fig1}}
\end{figure}

\clearpage

\begin{figure}
\plotone{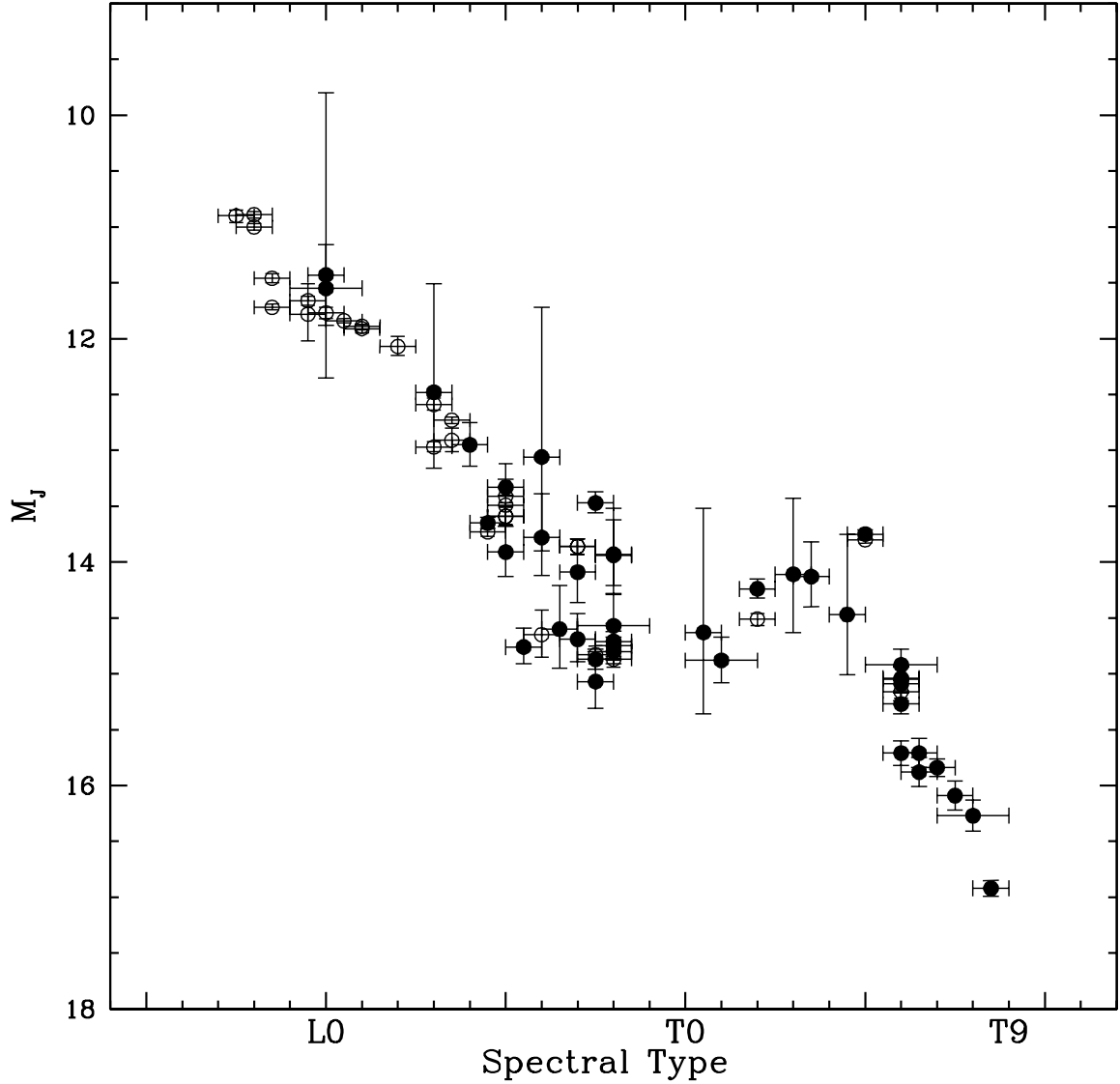}
\caption{ 
Absolute J-band magnitude ($M_J$) is plotted versus spectral type. The solid data points are the results from the infrared
astrometry and photometry from Table~6 of this paper; the open points are the results of optical astrometry from 
D02. For the seven objects in common, results are plotted for both optical and infrared parallaxes.
A complete description of this figure is given in \S 14.1.
\label{fig2}} \end{figure}

\clearpage

\begin{figure}
\plotone{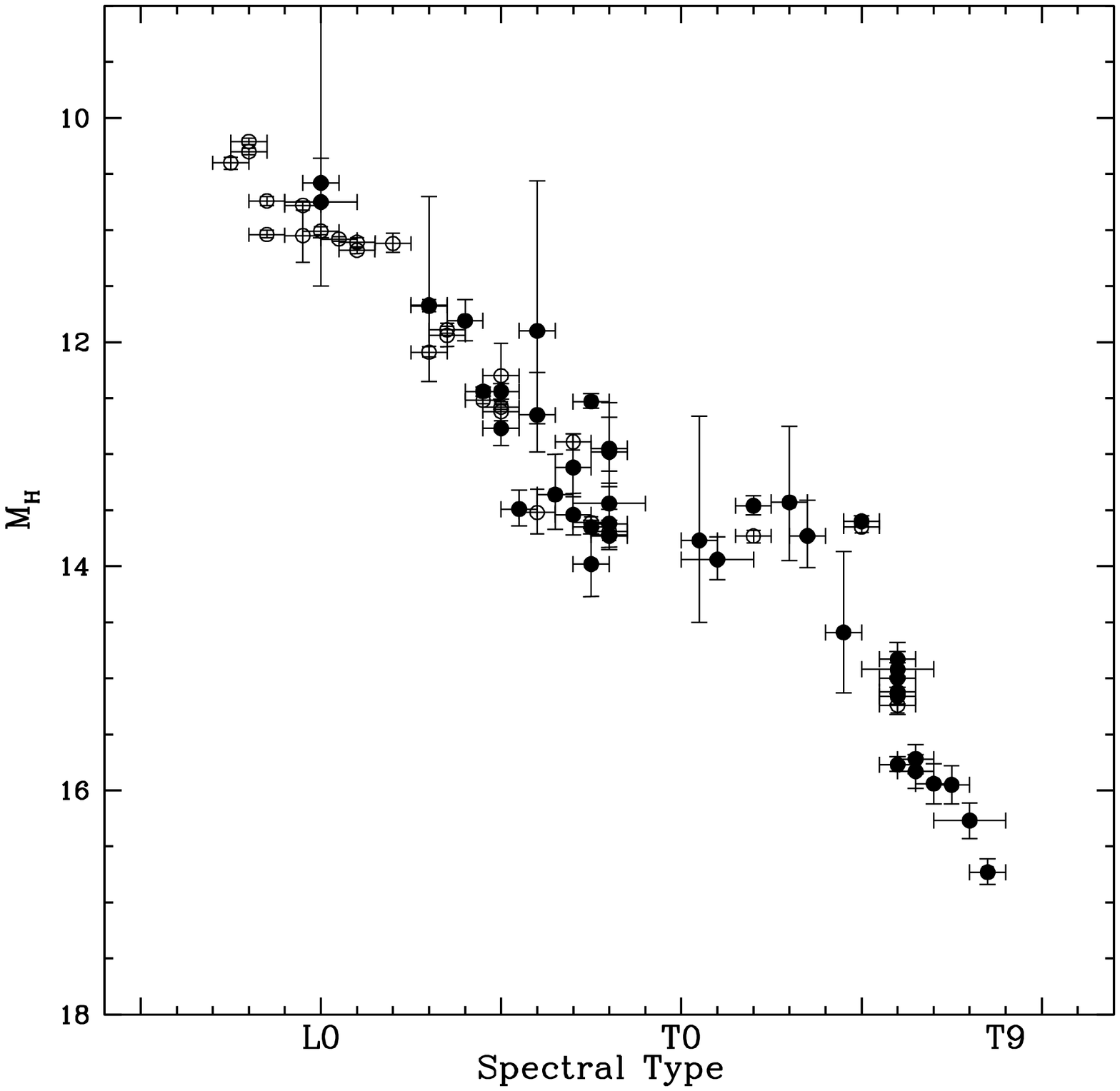}
\caption{Same as for Figure 2, except that absolute H-band magnitude ($M_H$) is plotted versus spectral type. \label{fig3}}
\end{figure}

\clearpage 

\begin{figure}
\plotone{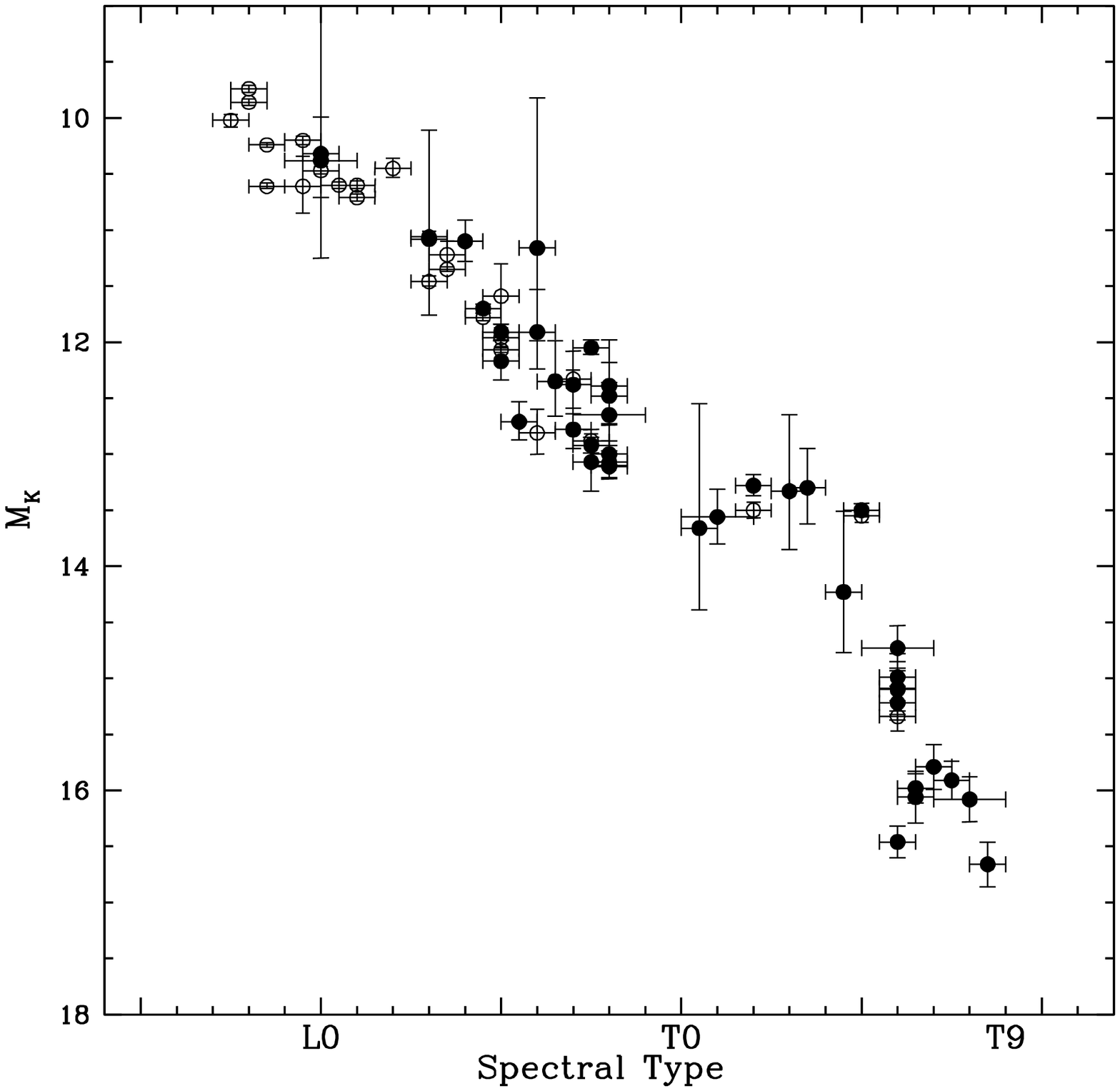}
\caption{Same as for Figure 2, except that absolute K-band magnitude ($M_K$) is plotted versus spectral type. \label{fig4}}
\end{figure}

\clearpage 

\begin{figure}
\plotone{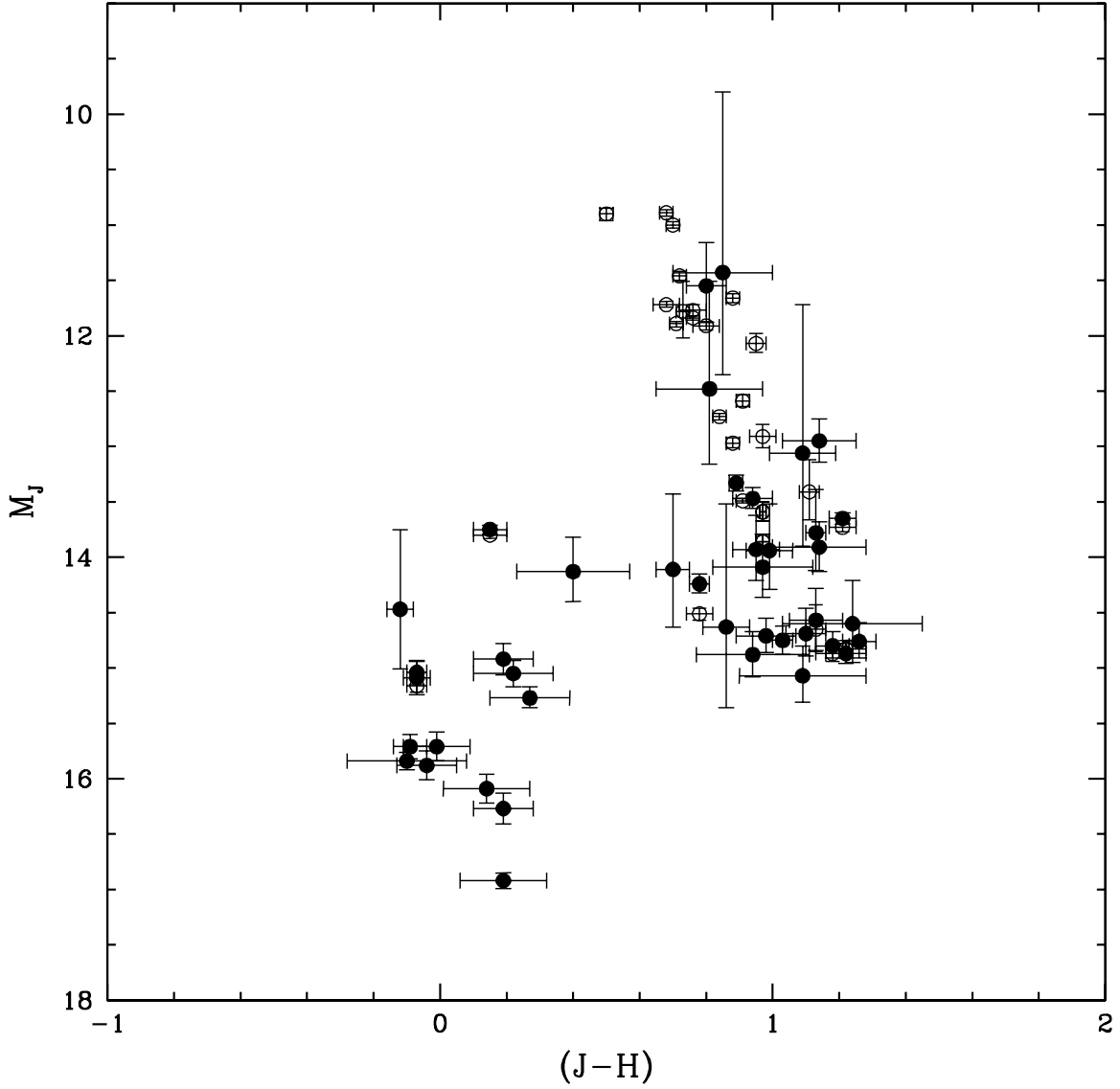}
\caption{Absolute J-band magnitude ($M_J$) is plotted versus ($J-H$) color. The solid data points are the results from the infrared
astrometry and photometry from Tables 2 and 6 of this paper, while the open points are the results of optical astrometry and
infrared photometry from D02. For the seven objects in common, results are plotted for both optical and infrared parallaxes.
A complete description of this figure is given in \S 14.2. \label{fig5}}
\end{figure}

\clearpage 

\begin{figure}
\plotone{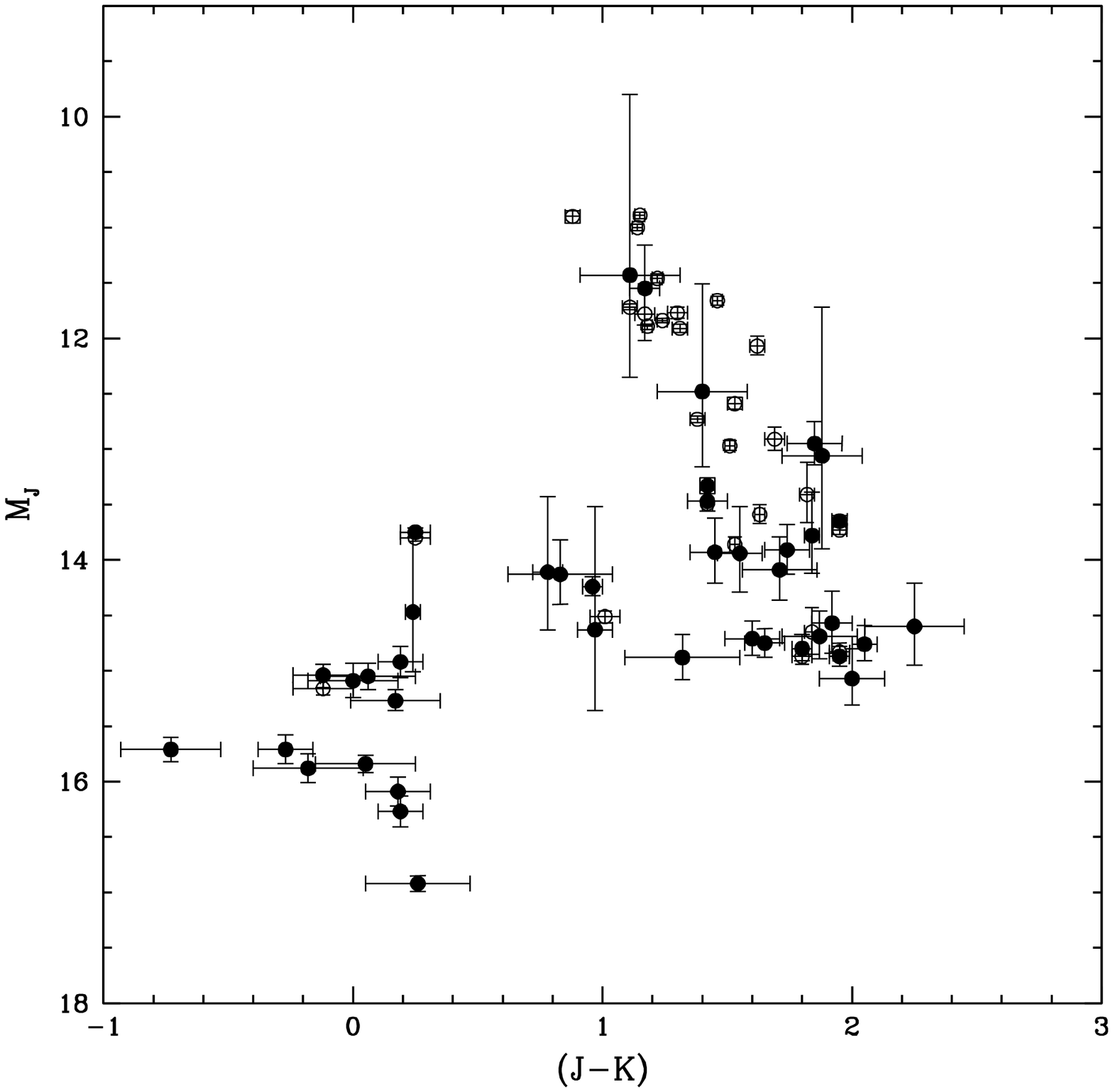}
\caption{Same as for Figure 5, except that $M_J$ is plotted versus ($J-K$) color.\label{fig6}}
\end{figure}

\clearpage 

\begin{figure}
\plotone{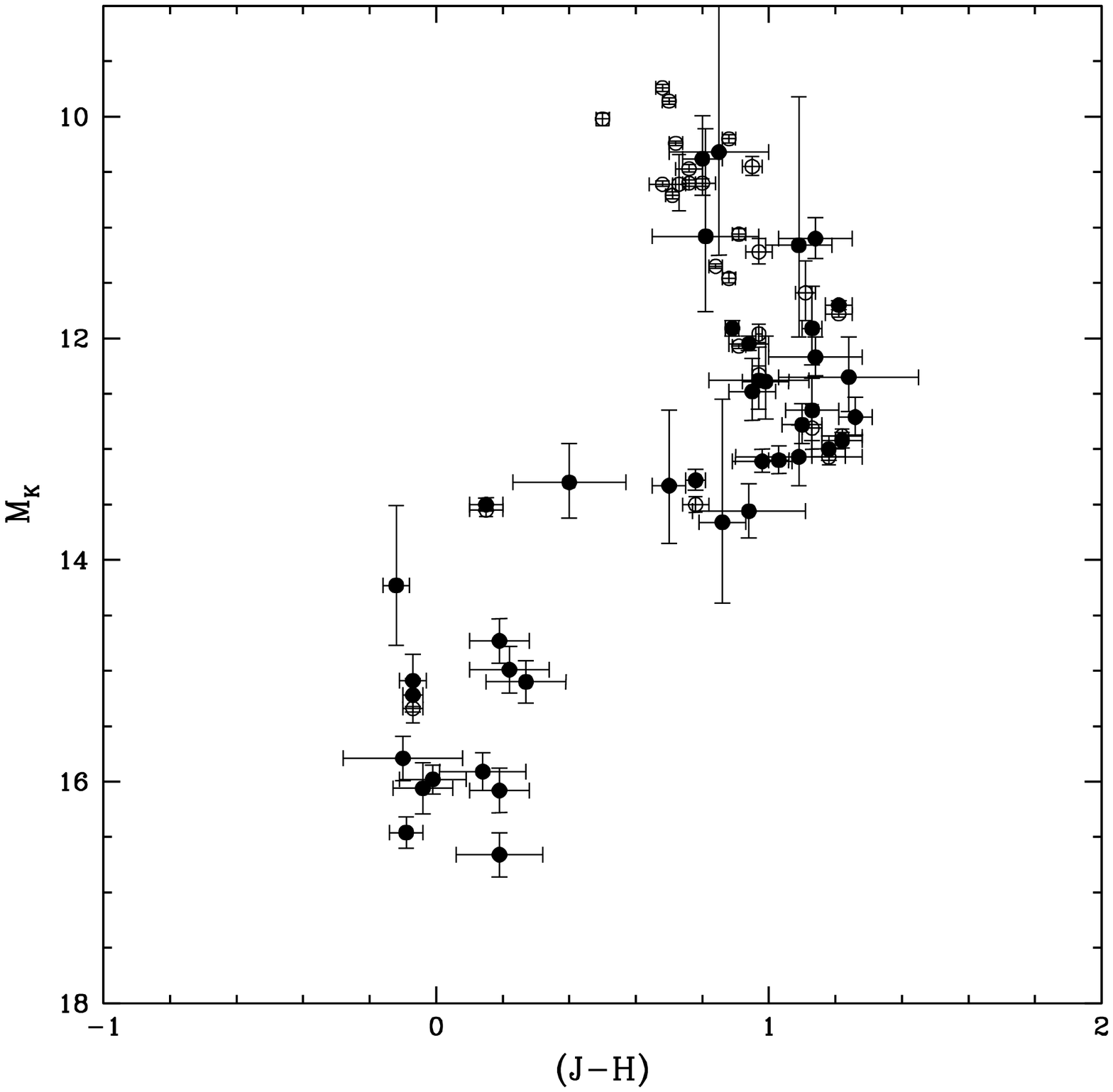}
\caption{Same as for Figure 5, except that $M_K$ is plotted versus ($J-H$) color. \label{fig7}}
\end{figure}

\clearpage 

\begin{figure}
\plotone{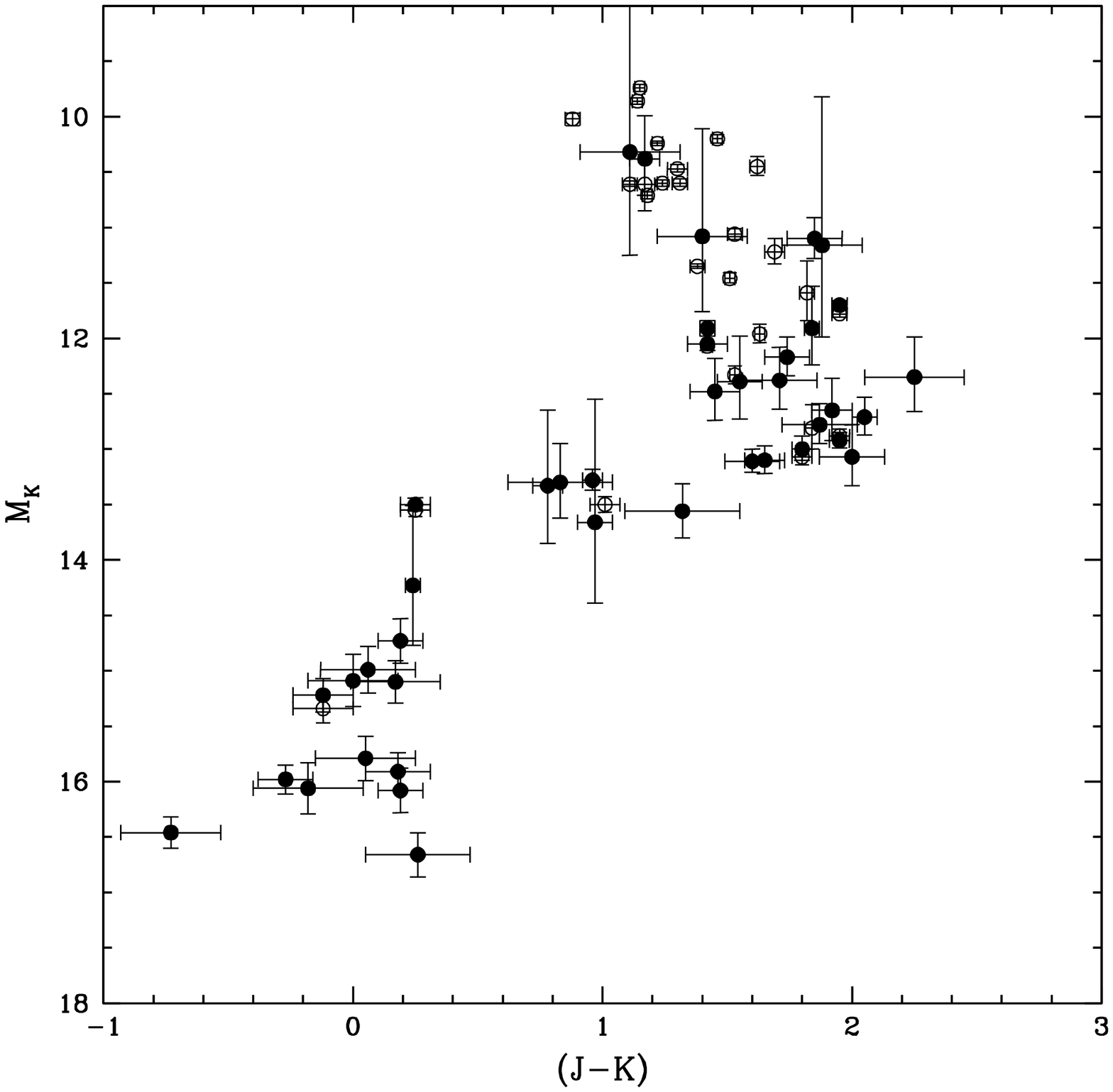}
\caption{Same as for Figure 5, except that $M_K$ is plotted versus ($J-K$) color. \label{fig8}}
\end{figure}

\clearpage 

\begin{figure}
\plotone{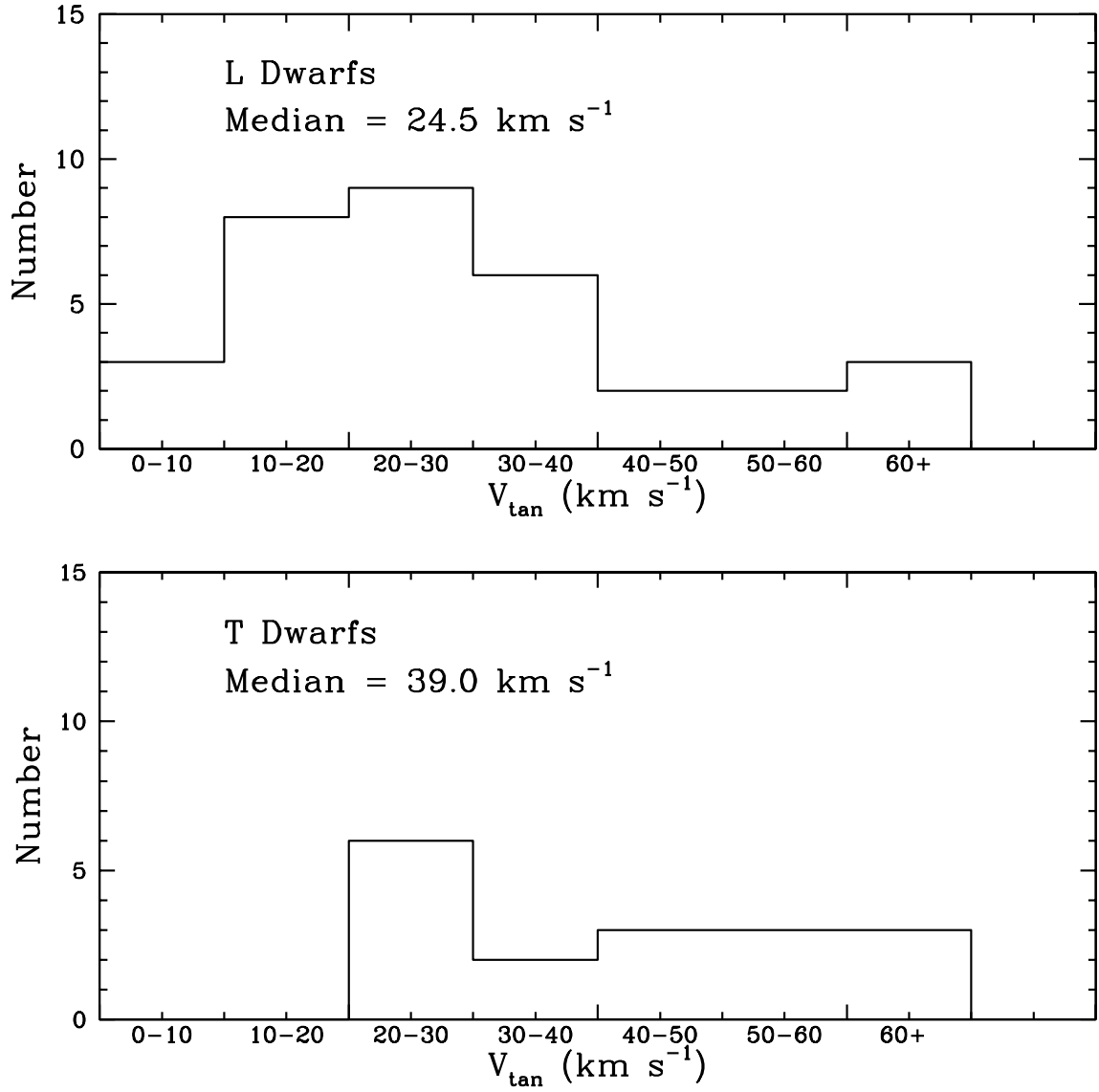}
\caption{Histograms of the distributions of V$_{tan}$ for L dwarfs and T dwarfs. \label{fig9}}
\end{figure}

\clearpage 

\begin{deluxetable}{llcccccc}
\tabletypesize{\scriptsize}
\tablecaption{Observations and Reference Stars. \label{tbl-1}}
\tablewidth{0pt}
\tablehead{
\colhead{Star} & \colhead{Sp.T.} & \colhead{Filt.}   & \colhead{No.  Nights}   &
\colhead{$\Delta$t (yrs)} & \colhead{Mean Epoch} & \colhead{No. Ref. Stars} & \colhead{Rel.$\rightarrow$Abs.} 
}
\startdata
2MASS J00303013$-$1450333    &  L7        & H & 31  &  2.18  &2001.800 & 10 & IR\\
SDSS J003259.37+141037.1     &  L8        & H & 23  &  2.02  &2002.159 & 14 & H,S,IR\\
SDSS J010752.42+004156.3     &  L5.5      & H & 34  &  1.87  &2002.308 &  6 & H,S\\
SDSS J015141.69+124429.6     &  T1$\pm$1  & H & 26  &  2.03  &2002.216 & 12 & H,S,IR\\
SDSS J020742.83+000056.2     &  T4.5      & H & 20  &  1.93  &2002.218 &  9 & H,S,IR\\
                             &            &   &     &        &          &    & \\    
2MASS J02431371$-$2453298    &  T6        & J & 26  &  2.20  &2001.774 &  5 & IR\\
2MASS J03284265+2302051      &  L8        & H & 22  &  2.19  &2001.745 &  9 & IR\\
2MASS J04151954$-$0935066    &  T8/T9      & J & 20  &  2.12  &2002.035 &  5 & IR\\
SDSS J042348.57$-$041403.5   &  L7.5/T0   & H & 21  &  2.02  &2002.248 &  9 & IR\\
SDSS J053952.00$-$005901.9   &  L5        & H & 22  &  2.02  &2001.970 & 10 & IR\\
                             &            &   &     &        &          &    & \\    
2MASS J05591914$-$1404488    &  T5        & J & 27  &  2.10  &2001.875 & 14 & IR\\
2MASS J07271824+1710012      &  T7        & J & 28  &  2.02  &2001.906 & 17 & IR\\
2MASS J08251968+2115521      &  L7.5      & H & 16  &  1.88  &2001.926 & 14 & IR\\
SDSS J083008.12+482847.4     &  L8        & H & 11  &  1.87  &2001.969 & 11 & H,S\\
SDSS J083717.21$-$000018.0   &  T0.5      & J & 10  &  1.78  &2002.016 & 12 & H,S,IR\\
                             &            &   &     &        &          &    & \\    
2MASS J08503593+1057156AB    &  L6+L/T    & H & 13  &  1.86  &2001.791 &  8 & IR\\
2MASS J09373487+2931409      &  T6p       & J & 14  &  1.87  &2001.984 &  6 & IR\\
2MASS J09510549+3558021      &  L6        & H & 15  &  1.28  &2001.725 &  9 & IR\\
SDSS J102109.69$-$030420.1   &  T3        & J & 15  &  1.31  &2001.743 &  8 & H,S,IR\\
2MASS J10475385+2124234      &  T6.5      & J & 22  &  1.32  &2001.739 &  7 & IR\\
                             &            &   &     &        &          &    & \\    
2MASS J12171110$-$0311131    &  T7.5      & J & 18  &  1.30  &2001.848 &  4 & H,S,IR\\
2MASS J12255432$-$2739466AB  &  T6:+T8:   & J & 14  &  1.30  &2001.919 & 12 & IR\\
2MASS J12373919+6526148      &  T6.5e     & J & 13  &  1.38  &2001.802 &  6 & H,S,IR\\
SDSS J125453.90$-$012247.4   &  T2        & J & 15  &  1.29  &2001.905 & 13 & H,S,IR\\
SDSS 132629.81$-$003831.4    &  L8?       & H & 17  &  1.38  &2001.829 & 10 & H,S,IR\\
                             &            &   &     &        &          &    & \\    
SDSS J134646.43$-$003150.4   &  T6        & J & 22  &  1.39  &2001.901 &  9 & H,S,IR\\
SDSS J143517.20$-$004612.9   &  L0        & H & 24  &  1.39  &2001.925 &  4 & H,S\\
SDSS J143535.72$-$004347.0   &  L3        & H & 24  &  1.39  &2001.923 &  6 & H,S\\
SDSS J144600.60+002452.0     &  L5        & H & 25  &  1.36  &2001.916 &  5 & H,S\\
2MASS J15232263+3014562      &  L8        & H & 32  &  1.36  &2001.888 & 10  & IR\\   
                             &            &   &     &        &          &    & \\    
SDSS J162414.36+002915.8     &  T6        & J & 24  &  1.37  &2001.920 &  6 & H,S\\
2MASS J16322911+1904407      &  L8        & H & 22  &  1.38  &2001.852 &  8 & IR\\
2MASS J17114573+2232044      &  L6.5      & H & 26  &  1.36  &2001.983 & 16 & H,S,IR\\
2MASS J17281150+3948593AB    &  L7+L/T    & H & 24  &  1.38  &2001.978 & 11 & IR\\
SDSS J175032.96+175903.9     &  T3.5      & J & 29  &  1.24  &2001.077 & 26 & H,S,IR\\
                             &            &   &     &        &          &    & \\    
2MASS J18410861+3117279      &  L4pec     & H & 39  &  2.03  &2002.048 & 19 & IR\\
2MASS J21011544+1756586AB    &  L7.5+L8?  & H & 48  &  2.18  &2002.055 &17 & IR\\
2MASS J22244381$-$0158521    &  L4.5      & H & 50  &  2.20  &2001.969 &6 & IR\\
SDSS J225529.09$-$003433.4   &  L0:       & H & 44  &  2.10  &2002.101 & 13 & H,S\\
2MASS J23565477$-$1553111    &  T6        & J & 43  &  2.20  &2001.939 & 10 & IR\\
\enddata
\end{deluxetable}

\clearpage 

\begin{deluxetable}{llccccc}
\tabletypesize{\scriptsize}
\tablecaption{Astrometric Results. \label{tbl-2}}
\tablewidth{0pt}
\tablehead{
\colhead{Star} & \colhead{Sp.T.} & \colhead{$\pi_{rel}$} & 
\colhead{$\pi_{abs}$}  & \colhead{$\mu_{rel}$} & 
\colhead{P.A.}  & \colhead{V$_{tan}$}  \\
\colhead{} & \colhead{} & \colhead{(mas)} & \colhead{(mas)} & \colhead{(mas
yr$^{-1}$)} & \colhead{(deg)} & \colhead{(km s$^{-1}$)}
}
\startdata
2MASS J003030$-$1450 &  L7   &  35.39 $\pm$ 4.49 &  37.42 $\pm$ 4.50 & 246.6  $\pm$ 3.6 
& 96.56 $\pm$ 0.42 & 31.2 $\pm$ 3.8\\
SDSS J003259+1410    &  L8   &  28.82 $\pm$ 5.16 &  30.14 $\pm$ 5.16 & 275.8  $\pm$ 6.9  
& 81.83 $\pm$ 0.72 & 43.4 $\pm$ 7.7\\ 
SDSS J010752+0041    &  L5.5 &  61.98 $\pm$ 4.48 &  64.13 $\pm$ 4.51 & 634.5  $\pm$ 7.1
& 81.71 $\pm$ 0.32 & 46.9 $\pm$ 3.3\\
SDSS J015141+1244    &  T1$\pm$1  &  45.26 $\pm$ 3.36 &  46.73 $\pm$ 3.37 & 742.7  $\pm$ 4.2   
& 92.84 $\pm$ 0.16 & 75.3 $\pm$ 5.5\\
SDSS J020742+0000    &  T4.5 &  33.15 $\pm$ 9.86 &  34.85 $\pm$ 9.87 & 156.3  $\pm$ 11.4     
& 96.29 $\pm$ 2.09 & 21.3 $\pm$ 6.7\\
               &          &   &     &        &    & \\    
2MASS J024313$-$2453 &  T6   & 90.23 $\pm$  3.50 &  93.62 $\pm$ 3.63 & 354.8  $\pm$ 4.1   
&234.20 $\pm$ 0.33 & 18.0 $\pm$ 0.7\\
2MASS J032842+2302   &  L8   & 31.06 $\pm$  4.19 &  33.13 $\pm$ 4.20 &  61.0  $\pm$ 4.9  
&168.08 $\pm$2.30 &  8.7 $\pm$ 1.3 \\
2MASS J041519$-$0935 & T8/T9   & 172.87 $\pm$ 2.75 & 174.34 $\pm$ 2.76 & 2255.3 $\pm$ 3.2 
& 76.48 $\pm$ 0.04 & 61.4 $\pm$ 1.0\\
SDSS J042348$-$0414  & L7.5/T0 &64.50 $\pm$ 1.69  &  65.93 $\pm$ 1.70 &  333.1 $\pm$ 2.8  
&284.23 $\pm$ 0.24 & 24.0 $\pm$ 0.6\\
SDSS J053952$-$0059 &  L5   &74.43 $\pm$ 2.16  &  76.12 $\pm$ 2.17 &  356.1 $\pm$ 3.5  
& 27.49 $\pm$ 0.28 & 22.2 $\pm$ 0.7\\
               &          &   &     &        &    & \\    
2MASS J055919$-$1404 &  T5   & 93.84 $\pm$ 1.43 &  95.53 $\pm$ 1.44  & 655.2 $\pm$ 2.8  
&121.10 $\pm$ 0.12 & 32.5 $\pm$ 0.5\\
2MASS J072718+1710 &  T7    &109.01 $\pm$ 2.34 & 110.14$\pm$  2.34  &1296.5 $\pm$ 4.5
&126.25 $\pm$ 0.10 & 55.8 $\pm$ 1.2\\
2MASS J082519+2115 &  L7.5 & 94.20 $\pm$ 1.83 &  95.64 $\pm$ 1.84  & 584.5 $\pm$ 4.0   
&239.45 $\pm$ 0.20 & 29.0 $\pm$ 0.6\\
SDSS J083008+4828 &  L8     & 74.93 $\pm$ 3.42 &  76.42 $\pm$ 3.43  &1267.0 $\pm$ 6.5    
&232.58 $\pm$ 0.15 & 78.6 $\pm$ 3.6\\
SDSS J083717$-$0000 &  T0.5  & 32.44 $\pm$ 13.45 & 33.70 $\pm$ 13.45 & 173.0 $\pm$ 16.7  
&185.05 $\pm$ 2.76 & 24.3 $\pm$ 11.8\\
               &          &   &     &        &    & \\    
2MASS J085035+1057AB &L6+L/T   & 24.76 $\pm$ 4.21 &  26.22 $\pm$ 4.21  & 147.2 $\pm$ 6.2   
&261.93 $\pm$ 1.20 & 26.6 $\pm$ 4.5\\
2MASS J093734+2931 &  T6p  &161.47 $\pm$ 3.87 & 162.84 $\pm$ 3.88  &1622.0 $\pm$ 7.1    
&143.14 $\pm$ 0.13 & 47.2 $\pm$ 1.1\\
2MASS J095105+3558 &  L6   & 14.88 $\pm$ 7.40 &  16.09 $\pm$ 7.40  & 189.4 $\pm$ 10.6   
&211.83 $\pm$ 1.60 & 55.8 $\pm$ 32.7\\
SDSS J102109$-$0304 &  T3   & 39.13 $\pm$ 11.00 & 40.78 $\pm$ 11.00 & 179.4 $\pm$ 15.2   
&246.24 $\pm$ 2.42 & 20.9 $\pm$ 6.3\\
2MASS J104753+2124 &  T6.5 & 92.82 $\pm$ 3.76 &  94.73 $\pm$ 3.81  &1728.4 $\pm$ 7.7    
&254.08 $\pm$ 0.13 & 86.5 $\pm$ 3.5\\
               &          &   &     &        &    & \\    
2MASS J121711$-$0311 &  T7.5 &108.63 $\pm$ 5.87 &  110.36 $\pm$ 5.88 &1061.8 $\pm$ 10.2    
&274.95 $\pm$ 0.27 & 45.6 $\pm$ 2.5\\
2MASS J122554$-$2739AB &T6:+T8: & 72.76 $\pm$ 3.46 & 74.20 $\pm$ 3.47  & 736.9 $\pm$ 6.8   
&147.80 $\pm$ 0.26 & 47.1 $\pm$ 2.3\\
2MASS J123739+6526 & T6.5e & 93.82 $\pm$ 4.75 & 96.07 $\pm$ 4.78  &1131.4 $\pm$ 8.9    
&242.33 $\pm$ 0.23 & 55.8 $\pm$ 2.8\\
SDSS J125453$-$0122 &  T2   & 74.51 $\pm$ 2.87 & 75.71 $\pm$ 2.88  & 483.2 $\pm$ 6.1   
&284.45 $\pm$ 0.36 & 30.3 $\pm$ 1.2\\
SDSS J132629$-$0038 &  L8? & 49.01 $\pm$ 6.33 & 49.98 $\pm$ 6.33  & 250.6 $\pm$ 8.9   
&244.63 $\pm$ 1.02 & 23.8 $\pm$ 3.2\\
               &          &   &     &        &    & \\    
SDSS J134646$-$0031 &  T6   & 71.13 $\pm$ 5.01 & 72.74 $\pm$ 5.02  & 491.5 $\pm$ 10.0   
&254.31 $\pm$ 0.58 & 32.0 $\pm$ 2.3\\
SDSS J143517$-$0046 &  L0   &  8.42 $\pm$ 5.10 &  9.85 $\pm$ 5.18  &  24.7 $\pm$ 9.2  
&64.99 $\pm$ 10.65 & 11.9 $\pm$ 9.7\\
SDSS J143535$-$0043 &  L3   & 15.28 $\pm$ 5.76 & 16.07 $\pm$ 5.76  & 107.5 $\pm$ 8.8   
&168.27 $\pm$ 2.34 & 31.7 $\pm$ 13.3\\
SDSS J144600+0024 &  L5   & 43.04 $\pm$ 3.21 & 45.46 $\pm$ 3.25  & 191.2 $\pm$ 7.0   
&110.06 $\pm$ 1.05 & 19.9 $\pm$ 1.6\\
2MASS J152322+3014 &  L8   & 55.77 $\pm$ 3.27  & 57.30 $\pm$ 3.27  & 221.4 $\pm$ 5.9
&139.77 $\pm$ 0.76 & 18.3 $\pm$ 1.2 \\   
               &          &   &     &        &    & \\    
SDSS J162414+0029 &  T6   & 84.94 $\pm$ 3.82 & 86.85 $\pm$ 3.85  & 374.0 $\pm$ 6.0   
&269.65 $\pm$ 0.46 & 20.4 $\pm$ 1.0\\
2MASS J163229+1904 &  L8   & 62.17 $\pm$ 3.31 & 63.58 $\pm$ 3.32  & 301.7 $\pm$ 5.1    
&101.37 $\pm$ 0.48 & 22.5 $\pm$ 1.2\\
2MASS J171145+2232 &  L6.5 & 31.48 $\pm$ 4.80 & 33.11 $\pm$ 4.81  &  31.2 $\pm$ 7.5  
& 98.35 $\pm$ 6.88 & 4.5 $\pm$ 1.2 \\
2MASS J172811+3948AB &  L7+L/T   & 40.33 $\pm$ 3.26 & 41.49 $\pm$ 3.26  &  45.0 $\pm$ 6.4  
&125.17 $\pm$ 4.07 &  5.1 $\pm$ 0.9\\
SDSS J175033+1759 & T3.5  & 35.07 $\pm$ 4.53 & 36.24 $\pm$ 4.53  & 204.3 $\pm$ 7.8   
& 60.80 $\pm$ 1.09 & 26.7 $\pm$ 3.5\\
               &          &   &     &        &    & \\    
2MASS J184108+3117 &  L4pec & 22.12 $\pm$ 1.88 & 23.57 $\pm$ 1.89  &  72.6 $\pm$ 3.7  
& 55.00 $\pm$ 1.46 & 14.6 $\pm$ 1.4\\
2MASS J210115+1756AB &  L7.5+L8? & 29.04 $\pm$ 3.42 & 30.14 $\pm$ 3.42  & 208.5 $\pm$ 3.7   
&136.33 $\pm$ 0.51 & 32.8 $\pm$ 3.8\\
2MASS J222443$-$0158 &  L4.5 & 83.43 $\pm$ 1.46 & 85.01 $\pm$ 1.50  & 980.6 $\pm$ 2.0  
&151.35 $\pm$ 0.06 & 54.7 $\pm$ 1.0\\
SDSS J225529$-$0034 &  L0:  & 14.61 $\pm$ 2.56 & 16.19 $\pm$ 2.59  & 179.9 $\pm$ 2.6   
&191.61 $\pm$ 0.41 & 52.7 $\pm$ 8.7\\
2MASS J235654$-$1553 &  T6   & 66.96 $\pm$ 3.38 & 68.97 $\pm$ 3.42  & 746.2 $\pm$ 2.9   
&216.46 $\pm$ 0.11 & 51.3 $\pm$ 2.6 \\
\enddata
\end{deluxetable}

\clearpage

\begin{deluxetable}{ccllll}
\tabletypesize{\scriptsize}
\tablecaption{Comparison of USNO Infrared and Optical Results. \label{tbl-3}}
\tablewidth{0pt}
\tablehead{
\colhead{Star} & \colhead{Program} & \colhead{$\Delta$t} & \colhead{$\pi_{abs}$} & 
\colhead{$\mu_{rel}$} & \colhead{P.A.} \\
\colhead{} & \colhead{} & \colhead{(yrs)} & \colhead{(mas)} & \colhead{(mas
yr$^{-1}$)} & \colhead{(deg)} 
}
\startdata
2MASS J055919$-$1404 & CCD\tablenotemark{1}    &  2.1  & 97.7  $\pm$ 1.3  & 661.2 $\pm$ 1.2 & 121.6  $\pm$ 0.1  \\
                 & IR     &  2.10 & 95.53 $\pm$ 1.44 & 655.2 $\pm$ 2.8 & 121.10 $\pm$ 0.12 \\
                 & CCD--IR &       & +2.2 $\pm$ 1.9 &   +6.0 $\pm$ 3.0  & $-$0.5 $\pm$ 0.2  \\   
                 &        &       &                  &                 &                   \\
2MASS J082519+2115   & CCD\tablenotemark{1}    &  2.0  & 93.8  $\pm$ 1.0  & 585.6 $\pm$ 1.4 & 240.5  $\pm$ 0.2  \\  
                 & IR     &  1.88 & 95.64 $\pm$ 1.84 & 584.5 $\pm$ 4.0 & 239.45 $\pm$ 0.20 \\
                 & CCD--IR &       &$-$1.8 $\pm$ 2.1 &   +1.1 $\pm$ 4.2  &  +1.0 $\pm$ 0.3  \\   
                 &        &       &                  &                 &                   \\
2MASS J085035+1057AB & CCD\tablenotemark{1}    &  3.3  & 39.1  $\pm$ 3.5  & 144.7 $\pm$ 2.0 & 267.0  $\pm$ 0.9  \\
                 & IR     &  1.86 & 26.22 $\pm$ 4.21 & 147.2 $\pm$ 6.2 & 261.93 $\pm$ 1.20 \\
                 & CCD--IR &       & +12.9 $\pm$ 5.5 & $-$2.5 $\pm$ 6.5  &   +5.1 $\pm$ 1.5 \\
                 &        &       &                  &                 &                   \\
SDSS J125453$-$0122 & CCD\tablenotemark{1}    &  1.2  & 84.9  $\pm$ 1.9  & 496.1 $\pm$ 1.8 & 285.2  $\pm$ 0.4  \\
                 & IR     &  1.29 & 75.71 $\pm$ 2.88 & 483.2 $\pm$ 6.1 & 284.45 $\pm$ 0.36 \\  
                 & CCD--IR &       & +9.2  $\pm$ 3.4 &  +12.9 $\pm$ 6.4  &   +0.7 $\pm$ 0.6 \\   
                 &        &       &                  &                 &                   \\
SDSS J162414+0029   & CCD\tablenotemark{1}    & 2.2   & 91.5  $\pm$ 2.3  & 383.2 $\pm$ 1.9 & 269.6  $\pm$ 0.5  \\ 
                 & IR     & 1.37  & 86.85 $\pm$ 3.85 & 374.0 $\pm$ 6.0 & 269.65 $\pm$ 0.46 \\
                 & CCD--IR &       &  +4.6 $\pm$ 4.5 & +9.2 $\pm$ 6.3  & $-$0.0 $\pm$ 0.7 \\   
                 &        &       &                  &                 &                   \\
2MASS J163229+1904   & CCD\tablenotemark{1}    &  3.2  & 65.6  $\pm$ 2.1  & 298.0 $\pm$ 0.9 & 100.4  $\pm$ 0.2  \\
                 & IR     &  1.38 & 63.58 $\pm$ 3.32 & 301.7 $\pm$ 5.1 & 101.37 $\pm$ 0.48 \\
                 & CCD--IR &       & +2.0  $\pm$ 3.9 & $-$3.7 $\pm$ 5.2  & $-$1.0 $\pm$ 0.5 \\    
                 &        &       &                  &                 &                   \\
2MASS J222443$-$0158 & CCD\tablenotemark{1}    & 2.3   & 88.1  $\pm$ 1.1  & 983.8 $\pm$ 0.7 & 152.3  $\pm$ 0.1  \\
                 & IR     & 2.20  & 85.01 $\pm$ 1.50 & 980.6 $\pm$ 2.0 & 151.35 $\pm$ 0.06 \\
                 & CCD--IR &       &  +3.1 $\pm$ 1.9 &   +3.8 $\pm$ 2.1  &   +0.9 $\pm$ 0.2 \\   
                 &        &       &                  &                 &                   \\
\cline{4-4} \cline{5-5} \cline{6-6}\\
Weighted Mean Diff.&      &       & +2.6 $\pm$ 1.4 &   +3.9 $\pm$ 1.5  & +0.3 $\pm$ 0.1  \\
\enddata
\tablenotetext{1}{D02}
\end{deluxetable}

\clearpage 

\begin{deluxetable}{ccllll}
\tabletypesize{\scriptsize}
\tablecaption{Comparison of Infrared Astrometric Results. \label{tbl-4}}
\tablewidth{0pt}
\tablehead{
\colhead{Star} & \colhead{Program} & \colhead{$\Delta$t} & \colhead{$\pi_{rel}$} & 
\colhead{$\mu_{rel}$} & \colhead{P.A.} \\
\colhead{} & \colhead{} & \colhead{(yrs)} & \colhead{(mas)} & \colhead{(mas
yr$^{-1}$)} & \colhead{(deg)} 
}
\startdata
SDSS J102109$-$0304   &  USNO  & 1.31 & 39.13 $\pm$ 11.00 & 179.4 $\pm$ 15.2 & 246.24 $\pm$ 2.42 \\
                   &  ESO\tablenotemark{1}   & 1.7  & 34.4  $\pm$  4.6  & 183.2 $\pm$  3.4 & 248.8  $\pm$ 1.0  \\
                   &  U--E   &      & +4.7  $\pm$ 11.9  &$-$3.8 $\pm$ 15.6 &$-$2.6  $\pm$ 2.6  \\
                   &        &      &                   &                  &                   \\
2MASS J104753+2124     &  USNO  & 1.32 & 92.82 $\pm$  3.76 & 1728.4 $\pm$ 7.7 & 254.08 $\pm$ 0.13 \\
                   &  ESO\tablenotemark{1}   & 1.9  &110.8  $\pm$  6.6  & 1698.9 $\pm$ 2.5 & 256.4  $\pm$ 0.1  \\ 
                   &  U--E   &     &$-$18.0 $\pm$  7.6  &  +29.5 $\pm$ 8.1 &$-$2.3  $\pm$ 0.2  \\
                   &        &      &                   &                  &                   \\
2MASS J121711$-$0311   &  USNO  & 1.30 &108.63 $\pm$  5.87 & 1061.8 $\pm$ 10.2& 274.95 $\pm$ 0.27 \\
                   &  ESO\tablenotemark{1}   & 1.9  & 90.8  $\pm$  2.2  & 1057.1 $\pm$ 1.7 & 274.1  $\pm$ 0.1  \\
                   &  U--E   &      &+17.8  $\pm$  6.3  &   +4.7 $\pm$ 10.3&  +0.9  $\pm$ 0.3  \\
                   &        &      &                   &                  &                   \\
2MASS J122554$-$2739AB &  USNO  & 1.30 & 72.76 $\pm$  3.46 &  736.9 $\pm$ 6.8 & 147.80 $\pm$ 0.26 \\
                   &  ESO\tablenotemark{1}   & 1.9  & 75.1  $\pm$  2.5  &  736.8 $\pm$ 2.9 & 148.5  $\pm$ 0.1  \\
                   &  U--E   &      &$-$2.3 $\pm$  4.3  &   +0.1 $\pm$ 7.4 &$-$0.7  $\pm$ 0.3  \\    
                   &        &      &                   &                  &                   \\
SDSS J125453$-$0122   &  USNO  & 1.29 & 74.51 $\pm$  2.87 &  483.2 $\pm$ 6.1 & 284.45 $\pm$ 0.36 \\
                   &  ESO\tablenotemark{1}   & 1.7  & 73.2  $\pm$  1.9  &  491.0 $\pm$ 2.5 & 284.7  $\pm$ 0.1  \\
                   &  U--E   &      & +1.3  $\pm$  3.4  & $-$7.8 $\pm$ 6.6 &$-$0.3  $\pm$ 0.4  \\
                   &        &      &                   &                  &                   \\
SDSS J134646$-$0031   &  USNO  & 1.39 & 71.13 $\pm$ 5.01  &  491.5 $\pm$ 10.0& 254.31 $\pm$ 0.58 \\
                   &  ESO\tablenotemark{1}   & 1.7  & 68.3  $\pm$ 2.3   &  516.0 $\pm$ 3.3 & 257.2  $\pm$ 0.2  \\  
                   &  U--E   &      & +2.8  $\pm$ 5.5   &$-$24.5 $\pm$ 10.5&$-$2.9  $\pm$ 0.6  \\
                   &        &      &                   &                  &                   \\
SDSS J162414+0029     &  USNO  & 1.37 & 84.94 $\pm$ 3.82  &  374.0 $\pm$ 6.0 & 269.65 $\pm$ 0.46 \\
                   &  ESO\tablenotemark{1}   & 1.9  & 90.9  $\pm$ 1.2   &  373.0 $\pm$ 1.6 & 268.6  $\pm$ 0.3  \\ 
                   &  U--E   &      &$-$6.0 $\pm$ 4.0   &   +1.0 $\pm$ 6.2 &   1.1  $\pm$ 0.5  \\
\cline{4-4} \cline{5-5} \cline{6-6}\\
Weighted Mean Diff.&        &      &$-$0.4 $\pm$ 3.1   &  +1.0  $\pm$ 5.7 &$-$1.0  $\pm$ 0.6  \\
(USNO-ESO)         &        &      &                   &                  &                   \\
\enddata
\tablenotetext{1}{TBK03}
\end{deluxetable}

\clearpage 

\begin{deluxetable}{lcccc}
\tabletypesize{\scriptsize}
\tablecaption{Additional USNO $JHK$ Photometry. \label{tbl-5}}
\tablewidth{0pt}
\tablehead{
\colhead{Star} & \colhead{Sp.T.} & \colhead{$K$ $\pm$ $\sigma$($K$)} & \colhead{$J-H$ $\pm$ $\sigma$($J-H$)} & 
\colhead{$H-K$ $\pm$ $\sigma$($H-K$)}  
}
\startdata
SDSS J020742+0000{\tablenotemark{*}}   &  T4.5  & 16.52 $\pm$ 0.03 & -0.12 $\pm$ 0.04 &  0.24 $\pm$ 0.03 \\
SDSS J053952$-$0059  &  L5   & 12.49 $\pm$ 0.04 &  0.89 $\pm$ 0.04 &  0.55 $\pm$ 0.05 \\  
2MASS J093734+2931   &  T6p  & 15.55 $\pm$ 0.15 & -0.12 $\pm$ 0.05 & -0.85 $\pm$ 0.15 \\    
2MASS J095105+3558   &  L6   & 15.10 $\pm$ 0.12 &  1.04 $\pm$ 0.13 &  0.77 $\pm$ 0.16 \\   
2MASS J104753+2124   &  T6.5 & 16.10 $\pm$ 0.10 & -0.12 $\pm$ 0.16 & -0.20 $\pm$ 0.16 \\    
                     &       &                  &                  &                  \\
SDSS J134646$-$0031  &  T6   & 15.76 $\pm$ 0.23 & -0.07 $\pm$ 0.04 &  0.12 $\pm$ 0.16 \\   
SDSS J162414+0029    &  T6   & 15.53 $\pm$ 0.12 & -0.08 $\pm$ 0.03 & -0.04 $\pm$ 0.12 \\   
2MASS J172811+3948AB &  L7+L/T   & 13.96 $\pm$ 0.11 &  1.06 $\pm$ 0.07 &  0.60 $\pm$ 0.10 \\  
2MASS J210115+1756AB &  L7.5+L8? & 16.90 $\pm$ 0.10 &  \nodata         &   \nodata        \\    
\enddata
\tablenotetext{*}{Based on relative photometry of 2MASS stars in the ASTROCAM field of view}
\end{deluxetable}

\clearpage

\begin{deluxetable}{lccrrrc}
\tabletypesize{\scriptsize}
\tablecaption{Adopted Spectral Types and $JHK$ Photometry. \label{tbl-6}}
\tablewidth{0pt}
\tablehead{
\colhead{Star} & \colhead{Sp.T.} & \colhead{Spec. Ref.} & \colhead{$K$ $\pm$ $\sigma$($K$)} & 
\colhead{($J-H$) $\pm$ $\sigma$($J-H$)} & \colhead{($J-K$) $\pm$ $\sigma$($J-K$)} & \colhead{Phot. Ref.} 
}
\startdata
2MASS J 003030$-$1450 &  L7       &  1  & 14.51 $\pm$ 0.10 &   0.97 $\pm$ 0.15  & 1.71 $\pm$ 0.15  & A    \\
SDSS J003259+1410     &  L8       &  3  & 14.99 $\pm$ 0.05 &   0.99 $\pm$ 0.07  & 1.55 $\pm$ 0.09  & A,B  \\
SDSS J010752+0041     &  L5.5     &  3  & 13.67 $\pm$ 0.07 &   1.26 $\pm$ 0.05  & 2.05 $\pm$ 0.05  & A,B  \\
SDSS J015141+1244     &  T1$\pm$1 &  3  & 15.21 $\pm$ 0.19 &   0.94 $\pm$ 0.17  & 1.32 $\pm$ 0.23  & A    \\
SDSS J020742+0000     &  T4.5     &  3  & 16.52 $\pm$ 0.03 &$-$0.12 $\pm$ 0.04  & 0.24 $\pm$ 0.03  & D    \\
                      &           &     &                  &                    &                  &      \\    
2MASS J024313$-$2453  &  T6       &  2  & 15.24 $\pm$ 0.17 &   0.27 $\pm$ 0.12  & 0.17 $\pm$ 0.18  & A    \\
2MASS J032842+2302    &  L8       &  1  & 14.88 $\pm$ 0.05 &   0.95 $\pm$ 0.07  & 1.45 $\pm$ 0.10  & A,B  \\
2MASS J041519$-$0935  &  T8/T9    & 2,4 & 15.45 $\pm$ 0.20 &   0.19 $\pm$ 0.13  & 0.26 $\pm$ 0.21  & A    \\
SDSS J042348$-$0414   &  L7.5/T0  &3,5,6& 12.95 $\pm$ 0.03 &   0.94 $\pm$ 0.06  & 1.42 $\pm$ 0.08  & A,B  \\
SDSS J053952$-$0059   &  L5       &  11 & 12.50 $\pm$ 0.04 &   0.89 $\pm$ 0.02  & 1.42 $\pm$ 0.03  & A,D,E\\
                      &           &     &                  &                    &                  &      \\ 
2MASS J055919$-$1404  &  T5       &  14 & 13.60 $\pm$ 0.05 &   0.15 $\pm$ 0.05  & 0.25 $\pm$ 0.06  & A    \\
2MASS J072718+1710    &  T7       &  2  & 15.58 $\pm$ 0.19 &$-$0.10 $\pm$ 0.18  & 0.05 $\pm$ 0.20  & A    \\
2MASS J082519+2115    &  L7.5     &  1  & 13.02 $\pm$ 0.05 &   1.22 $\pm$ 0.06  & 1.95 $\pm$ 0.04  & A,B,C\\
SDSS J083008+4828     &  L8       & 3,5 & 13.69 $\pm$ 0.03 &   0.98 $\pm$ 0.09  & 1.60 $\pm$ 0.11  & A,B  \\
SDSS J083717$-$0000   &  T0.5     &3,16 & 16.02 $\pm$ 0.05 &   0.86 $\pm$ 0.07  & 0.97 $\pm$ 0.07  & E    \\
                      &           &     &                  &                    &                  &      \\    
2MASS J085035+1057AB  &  L6+L/T   &13,15& 14.45 $\pm$ 0.04 &   1.13 $\pm$ 0.03  & 1.84 $\pm$ 0.03  & A,B,C\\
2MASS J093734+2931    &  T6p      &  2  & 15.40 $\pm$ 0.13 &$-$0.09 $\pm$ 0.05  &$-$0.73 $\pm$ 0.20& A,D  \\
2MASS J095105+3558    &  L6       &  1  & 15.13 $\pm$ 0.10 &   1.09 $\pm$ 0.10  & 1.88 $\pm$ 0.16  & A,D  \\
SDSS J102109$-$0304   &  T3       &3,16 & 15.28 $\pm$ 0.05 &   0.70 $\pm$ 0.05  & 0.78 $\pm$ 0.06  & A,E  \\
2MASS J104753+2124    &  T6.5     &  7  & 16.10 $\pm$ 0.10 &$-$0.01 $\pm$ 0.10  &$-$0.27 $\pm$ 0.11 & A,D  \\
                      &           &     &                  &                    &                  &      \\    
2MASS J121711$-$0311  &  T7.5     &  7  & 15.70 $\pm$ 0.12 &   0.14 $\pm$ 0.13  &   0.18 $\pm$ 0.13& A,C  \\
2MASS J122554$-$2739A &  T6:      & 7,10,12& 15.38 $\pm$ 0.17 &   0.19 $\pm$ 0.09  &   0.19 $\pm$ 0.16& A,F  \\
2MASS J122554$-$2739B &  T8:      & 7,10,12& 16.73 $\pm$ 0.17 &   0.19 $\pm$ 0.09  &   0.19 $\pm$ 0.16& A,F  \\
2MASS J123739+6526    &  T6.5e    &  7  & 16.15 $\pm$ 0.20 &$-$0.04 $\pm$ 0.09  &$-$0.18 $\pm$ 0.22& A,C  \\
SDSS J125453$-$0122   &  T2       &3,16 & 13.88 $\pm$ 0.04 &   0.78 $\pm$ 0.03  &   0.96 $\pm$ 0.04& A,E  \\
                      &           &     &                  &                    &                  &      \\    
SDSS J132629$-$0038   &  L8?      &  11 & 14.16 $\pm$ 0.05 &   1.13 $\pm$ 0.08  &   1.92 $\pm$ 0.08& A,B  \\
SDSS J134646$-$0031   &  T6       &3,19 & 15.78 $\pm$ 0.18 &$-$0.07 $\pm$ 0.04  &   0.00 $\pm$ 0.18& A,D  \\
SDSS J143517$-$0046   &  L0       &  8  & 15.35 $\pm$ 0.18 &   0.85 $\pm$ 0.15  &   1.11 $\pm$ 0.20& A    \\
SDSS J143535$-$0043   &  L3       &  8  & 15.05 $\pm$ 0.14 &   0.81 $\pm$ 0.16  &   1.40 $\pm$ 0.18& A    \\
SDSS J144600+0024     &  L5       & 2,3 & 13.88 $\pm$ 0.08 &   1.14 $\pm$ 0.14  &   1.74 $\pm$ 0.09& A,B  \\
                      &           &     &                  &                    &                  &      \\    
2MASS J152322+3014    &  L8       &  1  & 14.31 $\pm$ 0.03 &   1.03 $\pm$ 0.03  &   1.65 $\pm$ 0.08& A,B,C\\   
SDSS J162414+0029     &  T6       & 3,20& 15.53 $\pm$ 0.12 &$-$0.07 $\pm$ 0.03  &$-$0.12 $\pm$ 0.12& A,D  \\
2MASS J163229+1904    &  L8       &  13 & 13.98 $\pm$ 0.03 &   1.18 $\pm$ 0.05  &   1.80 $\pm$ 0.04& A,B,C\\
2MASS J171145+2232    &  L6.5     &  1  & 14.75 $\pm$ 0.10 &   1.24 $\pm$ 0.21  &   2.25 $\pm$ 0.20& A    \\
2MASS J172811+3948AB  &  L7+L/T   &1,17 & 13.94 $\pm$ 0.05 &   1.10 $\pm$ 0.06  &   1.87 $\pm$ 0.15& A,D  \\
                      &           &     &                  &                    &                  &      \\    
SDSS J175033+1759     &  T3.5     &  3  & 15.50 $\pm$ 0.19 &   0.40 $\pm$ 0.17  &   0.83 $\pm$ 0.21& A    \\
2MASS J184108+3117    &  L4pec    &  1  & 14.24 $\pm$ 0.07 &   1.14 $\pm$ 0.11  &   1.85 $\pm$ 0.11& A    \\
2MASS J210115+1756AB  &  L7.5+L8? &1,18 & 14.92 $\pm$ 0.12 &   1.09 $\pm$ 0.19  &   2.00 $\pm$ 0.13& A,D  \\
2MASS J222443$-$0158  &  L4.5     &  1  & 12.05 $\pm$ 0.02 &   1.21 $\pm$ 0.04  &   1.95 $\pm$ 0.03& A    \\
SDSS J225529$-$0034   &  L0:      &  9  & 14.33 $\pm$ 0.08 &   0.80 $\pm$ 0.06  &   1.17 $\pm$ 0.06& A,B  \\
2MASS J235654$-$1553  &  T6       &  2  & 15.80 $\pm$ 0.18 &   0.22 $\pm$ 0.12  &   0.06 $\pm$ 0.19& A    \\
\enddata
\tablerefs{(1) Kirkpatrick et al. (2000); (2) Burgasser et al. (2002a); (3) Geballe et al. (2002);
(4) Knapp et al. (2004); (5) Kirkpatrick (2004); (6) See \S 15.2; (7) Burgasser et al. (1999); 
(8) Hawley et al. (2002); (9) Schneider et al. (2002); (10) Burgasser et al. (2003);
(11) Fan et al. (2000); (12) See \S 15.6; (13) Kirkpatrick et al. (1999); (14) Burgasser et al. (2000);
(15) See \S 15.4; (16) Leggett et al. (2000); (17) See \S 15.8; (18) See \S 15.9; (19) Tsvetanov et al. (2000);
(20) Strauss et al. (1999); (A) 2MASS All-Sky Point Source Catalog; (B) Leggett et. al (2002); (C) D02;
(D) Table 5, this paper; (E) Leggett et. al (2000); (F) See \S 15.6}
\end{deluxetable}

\clearpage 

\begin{deluxetable}{llccccccc}
\tabletypesize{\scriptsize}
\tablecaption{Infrared Absolute Magnitudes based on USNO Astrometry. \label{tbl-7}}
\tablewidth{0pt}
\tablehead{
\colhead{Star} & \colhead{Sp.T.} & \colhead{Opt/IR} & \colhead{$M_J$} & \colhead {$\sigma$($-$) $\sigma$(+)} & 
\colhead{$M_H$} & \colhead {$\sigma$($-$) $\sigma$(+)} & 
\colhead{$M_K$} & \colhead {$\sigma$($-$) $\sigma$(+)}  
}
\startdata
2MASS J034543+2540     &  L0         &  O & 11.77 & .05 .05 &  11.01 & .04 .03 &  10.47 &  .04 .03 \\   
SDSS J143517$-$0046    &  L0         &  I & 11.43 &1.63 .92 &  10.58 &1.63 .92 &  10.32 & 1.63 .93 \\
SDSS J225529$-$0034    &  L0:        &  I & 11.55 & .39 .33 &  10.75 & .39 .32 &  10.38 &  .39 .33 \\
2MASS J074642+2000A    &  L0.5       &  O & 11.84 & .02 .02 &  11.08 & .02 .02 &  10.60 &  .03 .03 \\
2MASS J143928+1929     &  L1         &  O & 11.89 & .02 .03 &  11.18 & .02 .03 &  10.71 &  .03 .03 \\
2MASS J165803+7027     &  L1         &  O & 11.91 & .04 .04 &  11.11 & .04 .04 &  10.60 &  .04 .03 \\
Kelu$-$1               &  L2         &  O & 12.07 & .09 .08 &  11.12 & .09 .08 &  10.45 &  .09 .08 \\
DENIS J1058.7$-$1548   &  L3         &  O & 12.97 & .05 .04 &  12.09 & .05 .04 &  11.46 &  .05 .04 \\
2MASS J114634+2230A    &  L3         &  O & 12.59 & .06 .05 &  11.68 & .06 .05 &  11.06 &  .05 .05 \\
2MASS J143535$-$0043   &  L3         &  I & 12.48 & .97 .68 &  11.67 & .97 .68 &  11.08 &  .97 .68 \\
2MASS J003616+1821     &  L3.5       &  O & 12.73 & .03 .03 &  11.89 & .03 .03 &  11.35 &  .02 .02 \\
2MASS J032613+2950     &  L3.5       &  O & 12.91 & .11 .10 &  11.94 & .11 .10 &  11.22 &  .12 .11 \\ 
2MASS J184108+3117     & L4pec       &  I & 12.95 & .20 .19 &  11.81 & .19 .18 &  11.10 &  .19 .18 \\ 
2MASS J222443$-$0158   &  L4.5       &I+O & 13.70 & .05 .05 &  12.49 & .04 .04 &  11.75 &  .04 .04 \\
SDSS J053952$-$0059    &  L5         &  I & 13.33 & .07 .07 &  12.44 & .07 .07 &  11.91 &  .07 .07 \\
DENIS J122815$-$1547A  &  L5         &  O & 13.59 & .09 .08 &  12.62 & .09 .08 &  11.96 &  .09 .08 \\
DENIS J122815$-$1547B  &  L5         &  O & 13.59 & .09 .08 &  12.62 & .09 .08 &  11.96 &  .09 .08 \\
2MASS J132855+2114     &  L5         &  O & 13.41 & .29 .25 &  12.30 & .29 .25 &  11.59 &  .29 .25 \\
SDSS J144600+0024      &  L5         &  I & 13.91 & .23 .22 &  12.77 & .17 .15 &  12.17 &  .18 .17 \\
2MASS J150747$-$1627   &  L5         &  O & 13.49 & .02 .02 &  12.58 & .02 .02 &  12.07 &  .02 .02 \\
SDSS J010752+0041      &  L5.5       &  I & 14.76 & .17 .15 &  13.49 & .17 .15 &  12.71 &  .18 .16 \\
2MASS J085035+1057A    &  L6         &I+O & 14.34 & .46 .38 &  13.21 & .45 .37 &  12.50 &  .45 .37 \\
2MASS J095105+3558     &  L6         &  I & 13.06 &1.34 .84 &  11.90 &1.34 .83 &  11.16 & 1.34 .83 \\
2MASS J171145+2232     &  L6.5       &  I & 14.60 & .39 .35 &  13.36 & .36 .31 &  12.35 &  .36 .31 \\ 
2MASS J003030$-$1450   &  L7         &  I & 14.09 & .30 .27 &  13.12 & .30 .26 &  12.38 &  .30 .26 \\ 
2MASS J172811+3948A    &  L7         &  I & 14.69 & .23 .20 &  13.54 & .19 .18 &  12.78 &  .19 .17 \\
DENIS J020529$-$1159A  &  L7         &  O & 13.86 & .07 .07 &  12.89 & .07 .07 &  12.33 &  .08 .08 \\
DENIS J020529$-$1159B  &  L7         &  O & 13.86 & .07 .07 &  12.89 & .07 .07 &  12.33 &  .08 .08 \\
SDSS J042348$-$0414    &  L7.5/T0    &  I & 13.47 & .10 .09 &  12.53 & .07 .06 &  12.05 &  .07 .06 \\
2MASS J082519+2115     &  L7.5       &I+O & 14.84 & .08 .08 &  13.62 & .04 .04 &  12.89 &  .05 .05 \\
2MASS J210115+1756A    &  L7.5       &  I & 15.07 & .27 .24 &  13.98 & .32 .29 &  13.07 &  .29 .26 \\
SDSS J003259+1410      &  L8         &  I & 13.94 & .42 .35 &  12.95 & .41 .34 &  12.39 &  .41 .34 \\
2MASS J032842+2302     &  L8         &  I & 13.93 & .31 .28 &  12.98 & .31 .28 &  12.48 &  .30 .26 \\
SDSS J083008+4828      &  L8         &  I & 14.71 & .16 .15 &  13.73 & .11 .10 &  13.11 &  .11 .10 \\
SDSS J132629$-$0038    &  L8?        &  I & 14.57 & .29 .27 &  13.44 & .29 .27 &  12.65 &  .29 .27 \\
2MASS J152322+3014     &  L8         &  I & 14.75 & .13 .13 &  13.72 & .13 .13 &  13.10 &  .13 .12 \\
2MASS J163229+1904     &  L8         &I+O & 14.85 & .08 .07 &  13.67 & .08 .07 &  13.05 &  .07 .06 \\
SDSS J083717$-$0000    &  T0.5       &  I & 14.63 &1.11 .73 &  13.77 &1.11 .73 &  13.66 & 1.11 .73 \\
SDSS J015141+1244      &  T1$\pm$1   &  I & 14.88 & .21 .20 &  13.94 & .20 .18 &  13.56 &  .25 .24 \\  
SDSS J125453$-$0122    &  T2         &I+O & 14.41 & .12 .12 &  13.63 & .12 .12 &  13.45 &  .12 .12 \\
SDSS J102109$-$0304    &  T3         &  I & 14.11 & .68 .52 &  13.43 & .68 .52 &  13.33 &  .68 .52 \\
SDSS J175033+1759      &  T3.5       &  I & 14.13 & .31 .27 &  13.73 & .32 .28 &  13.30 &  .35 .32 \\
SDSS J020742+0000      &  T4.5       &  I & 14.47 & .72 .54 &  14.59 & .72 .54 &  14.23 &  .72 .54 \\
2MASS J055919$-$1404   &  T5         &I+O & 13.78 & .03 .03 &  13.63 & .04 .04 &  13.53 &  .05 .06 \\
2MASS J024313$-$2453   &  T6         &  I & 15.27 & .10 .09 &  15.00 & .14 .14 &  15.10 &  .19 .19 \\
2MASS J093734+2931     &  T6p        &  I & 15.71 & .11 .11 &  15.77 & .07 .06 &  16.46 &  .14 .14 \\
2MASS J122554$-$2739A  &  T6:        &  I & 14.92 & .14 .14 &  14.92 & .16 .16 &  14.73 &  .20 .20 \\
SDSS J134646$-$0031    &  T6         &  I & 15.09 & .16 .15 &  15.16 & .17 .16 &  15.09 &  .24 .23 \\
SDSS J162414+0029      &  T6         &I+O & 15.13 & .06 .05 &  15.21 & .07 .06 &  15.31 &  .13 .13 \\
2MASS J235654$-$1553   &  T6         &  I & 15.05 & .12 .12 &  14.83 & .15 .15 &  14.99 &  .21 .21 \\
2MASS J104753+2124     &  T6.5       &  I & 15.71 & .13 .13 &  15.72 & .13 .13 &  15.98 &  .13 .13 \\
2MASS J123739+6526     &  T6.5e      &  I & 15.88 & .13 .13 &  15.83 & .15 .15 &  16.06 &  .23 .23 \\    
2MASS J072718+1710     &  T7         &  I & 15.84 & .08 .08 &  15.94 & .18 .18 &  15.79 &  .20 .20 \\ 
2MASS J121711$-$0311   &  T7.5       &  I & 16.09 & .13 .13 &  15.95 & .17 .17 &  15.91 &  .17 .17 \\
2MASS J122554$-$2739B  &  T8:        &  I & 16.27 & .14 .14 &  16.27 & .16 .16 &  16.08 &  .20 .20 \\
2MASS J041519$-$0935   &  T8/T9      &  I & 16.92 & .07 .07 &  16.73 & .12 .11 &  16.66 &  .20 .20 \\
\enddata
\end{deluxetable}

\clearpage 

\begin{deluxetable}{llccc}
\tabletypesize{\scriptsize}
\tablecaption{Derived L and T Dwarf Bolometric Magnitudes, Luminosities, and T$_{\rm eff}$. \label{tbl-8}}
\tablewidth{0pt}
\tablehead{
\colhead{Star} & \colhead{Sp.T.} &  \colhead{$M_{bol}$ $\pm$ $\sigma$($M_{bol}$)} & 
\colhead{log(L/L$_{\odot}$) $\pm$ $\sigma$[log(L/L$_{\odot}$)]} &  
\colhead{T$_{\rm eff}$ $\pm$ $\sigma$(T$_{\rm eff}$) (K)}  
}
\startdata
2MASS J034543+2540    &  L0         & 13.68 $\pm$ 0.14   &  $-$3.58 $\pm$ 0.06   &  2426 $^{+246}_{-191}$  \\   
SDSS J143517$-$0046   &  L0         & 13.53 $\pm$ 1.29   &  $-$3.52 $\pm$ 0.52   &  2511 $^{+915}_{-664}$  \\
SDSS J225529$-$0034   &  L0:        & 13.59 $\pm$ 0.39   &  $-$3.54 $\pm$ 0.16   &  2477 $^{+338}_{-276}$  \\
2MASS J074642+2000A   &  L0.5       & 13.84 $\pm$ 0.14   &  $-$3.64 $\pm$ 0.06   &  2338 $^{+238}_{-187}$  \\
2MASS J143928+1929    &  L1         & 13.98 $\pm$ 0.14   &  $-$3.70 $\pm$ 0.06   &  2264 $^{+230}_{-181}$  \\
2MASS J165803+7027    &  L1         & 13.87 $\pm$ 0.14   &  $-$3.65 $\pm$ 0.06   &  2322 $^{+236}_{-186}$  \\
Kelu$-$1              &  L2         & 13.76 $\pm$ 0.16   &  $-$3.61 $\pm$ 0.06   &  2382 $^{+246}_{-195}$  \\
DENIS J1058.7$-$1548  &  L3         & 14.79 $\pm$ 0.14   &  $-$4.02 $\pm$ 0.06   &  1879 $^{+191}_{-150}$  \\
2MASS J114634+2230A   &  L3         & 14.39 $\pm$ 0.14   &  $-$3.86 $\pm$ 0.06   &  2060 $^{+209}_{-165}$  \\
2MASS J143535$-$0043  &  L3         & 14.41 $\pm$ 0.84   &  $-$3.87 $\pm$ 0.34   &  2051 $^{+489}_{-386}$  \\
2MASS J003616+1821    &  L3.5       & 14.69 $\pm$ 0.13   &  $-$3.98 $\pm$ 0.05   &  1923 $^{+193}_{-153}$  \\
2MASS J032613+2950    &  L3.5       & 14.56 $\pm$ 0.18   &  $-$3.93 $\pm$ 0.07   &  1981 $^{+209}_{-166}$  \\ 
2MASS J184108+3117    & L4pec       & 14.45 $\pm$ 0.23   &  $-$3.88 $\pm$ 0.09   &  2032 $^{+226}_{-181}$  \\ 
2MASS J222443$-$0158  &  L4.5       & 15.09 $\pm$ 0.14   &  $-$4.14 $\pm$ 0.06   &  1753 $^{+179}_{-140}$  \\
SDSS J053952$-$0059   &  L5         & 15.25 $\pm$ 0.15   &  $-$4.20 $\pm$ 0.06   &  1690 $^{+173}_{-137}$  \\
DENIS J122815$-$1547A &  L5         & 15.30 $\pm$ 0.16   &  $-$4.22 $\pm$ 0.06   &  1671 $^{+172}_{-137}$  \\
DENIS J122815$-$1547B &  L5         & 15.30 $\pm$ 0.16   &  $-$4.22 $\pm$ 0.06   &  1671 $^{+172}_{-137}$  \\
2MASS J132855+2114    &  L5         & 14.93 $\pm$ 0.30   &  $-$4.08 $\pm$ 0.12   &  1819 $^{+221}_{-178}$  \\
SDSS J144600+0024     &  L5         & 15.51 $\pm$ 0.22   &  $-$4.31 $\pm$ 0.09   &  1592 $^{+175}_{-140}$  \\
2MASS J150747$-$1627  &  L5         & 15.41 $\pm$ 0.13   &  $-$4.27 $\pm$ 0.05   &  1629 $^{+164}_{-129}$  \\
SDSS J010752+0041     &  L5.5       & 16.04 $\pm$ 0.21   &  $-$4.52 $\pm$ 0.08   &  1409 $^{+153}_{-122}$  \\
2MASS J085035+1057A   &  L6         & 15.81 $\pm$ 0.43   &  $-$4.43 $\pm$ 0.17   &  1486 $^{+214}_{-175}$  \\
2MASS J095105+3558    &  L6         & 14.47 $\pm$ 1.10   &  $-$3.89 $\pm$ 0.44   &  2023 $^{+624}_{-471}$  \\
2MASS J171145+2232    &  L6.5       & 15.64 $\pm$ 0.36   &  $-$4.36 $\pm$ 0.14   &  1545 $^{+203}_{-165}$  \\ 
2MASS J003030$-$1450  &  L7         & 15.64 $\pm$ 0.31   &  $-$4.36 $\pm$ 0.12   &  1545 $^{+190}_{-154}$  \\ 
2MASS J172811+3948A   &  L7         & 16.04 $\pm$ 0.22   &  $-$4.52 $\pm$ 0.09   &  1409 $^{+155}_{-124}$  \\
DENIS J020529$-$1159A &  L7         & 15.59 $\pm$ 0.15   &  $-$4.34 $\pm$ 0.06   &  1563 $^{+160}_{-127}$  \\
DENIS J020529$-$1159B &  L7         & 15.59 $\pm$ 0.15   &  $-$4.34 $\pm$ 0.06   &  1563 $^{+160}_{-127}$  \\
SDSS J042348$-$0414   &  L7.5/T0    & 15.28 $\pm$ 0.16   &  $-$4.22 $\pm$ 0.06   &  1678 $^{+174}_{-137}$  \\
2MASS J082519+2115    &  L7.5       & 16.12 $\pm$ 0.15   &  $-$4.55 $\pm$ 0.06   &  1383 $^{+142}_{-112}$  \\
2MASS J210115+1756A   &  L7.5       & 16.30 $\pm$ 0.31   &  $-$4.62 $\pm$ 0.12   &  1327 $^{+163}_{-132}$  \\
SDSS J003259+1410     &  L8         & 15.58 $\pm$ 0.41   &  $-$4.34 $\pm$ 0.16   &  1566 $^{+220}_{-179}$  \\
2MASS J032842+2302    &  L8         & 15.67 $\pm$ 0.31   &  $-$4.37 $\pm$ 0.12   &  1534 $^{+189}_{-153}$  \\
SDSS J083008+4828     &  L8         & 16.30 $\pm$ 0.18   &  $-$4.62 $\pm$ 0.07   &  1327 $^{+140}_{-111}$  \\
SDSS J132629$-$0038   &  L8?        & 15.84 $\pm$ 0.32   &  $-$4.44 $\pm$ 0.13   &  1475 $^{+184}_{-149}$  \\
2MASS J152322+3014    &  L8         & 16.29 $\pm$ 0.19   &  $-$4.62 $\pm$ 0.08   &  1330 $^{+142}_{-112}$  \\
2MASS J163229+1904    &  L8         & 16.24 $\pm$ 0.16   &  $-$4.60 $\pm$ 0.06   &  1346 $^{+139}_{-110}$  \\
SDSS J083717$-$0000   &  T0.5       & 16.59 $\pm$ 0.93   &  $-$4.74 $\pm$ 0.37   &  1241 $^{+326}_{-252}$  \\
SDSS J015141+1244     &  T1$\pm$1   & 16.43 $\pm$ 0.31   &  $-$4.68 $\pm$ 0.12   &  1288 $^{+158}_{-128}$  \\  
SDSS J125453$-$0122   &  T2         & 16.19 $\pm$ 0.19   &  $-$4.58 $\pm$ 0.08   &  1361 $^{+145}_{-115}$  \\
SDSS J102109$-$0304   &  T3         & 15.93 $\pm$ 0.62   &  $-$4.48 $\pm$ 0.25   &  1445 $^{+267}_{-216}$  \\
SDSS J175033+1759     &  T3.5       & 15.83 $\pm$ 0.37   &  $-$4.44 $\pm$ 0.15   &  1478 $^{+197}_{-160}$  \\
SDSS J020742+0000     &  T4.5       & 16.63 $\pm$ 0.65   &  $-$4.76 $\pm$ 0.26   &  1230 $^{+236}_{-191}$  \\
2MASS J055919$-$1404  &  T5         & 15.86 $\pm$ 0.17   &  $-$4.45 $\pm$ 0.06   &  1469 $^{+153}_{-122}$  \\
2MASS J024313$-$2453  &  T6         & 17.31 $\pm$ 0.25   &  $-$5.03 $\pm$ 0.10   &  1052 $^{+119}_{-97}$   \\
2MASS J093734+2931    &  T6p        & 18.67 $\pm$ 0.21   &  $-$5.57 $\pm$ 0.08   &   769 $^{+85}_{-69}$    \\
2MASS J122554$-$2739A &  T6:        & 16.94 $\pm$ 0.27   &  $-$4.88 $\pm$ 0.11   &  1145 $^{+133}_{-188}$  \\
SDSS J134646$-$0031   &  T6         & 17.30 $\pm$ 0.29   &  $-$5.02 $\pm$ 0.12   &  1054 $^{+126}_{-102}$  \\
SDSS J162414+0029     &  T6         & 17.52 $\pm$ 0.20   &  $-$5.11 $\pm$ 0.08   &  1002 $^{+98}_{-86}$    \\
2MASS J235654$-$1553  &  T6         & 17.20 $\pm$ 0.26   &  $-$4.98 $\pm$ 0.10   &  1079 $^{+124}_{-100}$  \\
2MASS J104753+2124    &  T6.5       & 18.14 $\pm$ 0.20   &  $-$5.36 $\pm$ 0.08   &   869 $^{+93}_{-75}$    \\
2MASS J123739+6526    &  T6.5e      & 18.22 $\pm$ 0.28   &  $-$5.39 $\pm$ 0.11   &   853 $^{+101}_{-82}$   \\    
2MASS J072718+1710    &  T7         & 17.90 $\pm$ 0.25   &  $-$5.26 $\pm$ 0.10   &   918 $^{+105}_{-84}$   \\ 
2MASS J121711$-$0311  &  T7.5       & 17.98 $\pm$ 0.23   &  $-$5.30 $\pm$ 0.09   &   901 $^{+101}_{-80}$   \\
2MASS J122554$-$2739B &  T8:        & 18.12 $\pm$ 0.27   &  $-$5.35 $\pm$ 0.11   &   873 $^{+102}_{-82}$   \\
2MASS J041519$-$0935  &  T8/T9      & 18.70 $\pm$ 0.26   &  $-$5.58 $\pm$ 0.10   &   764 $^{+88}_{-71}$    \\
\enddata
\end{deluxetable}

\clearpage 

\begin{deluxetable}{lccccrccc}
\tabletypesize{\scriptsize}
\tablecaption{Mean L and T Dwarf Characteristics. \label{tbl-9}}
\tablewidth{0pt}
\tablehead{
\colhead{Sp.T.} & \colhead{$M_J$} &  \colhead{$M_H$} & \colhead{$M_K$} & \colhead{$J-H$} & \colhead{$J-K$} &
\colhead{$M_{bol}$} & \colhead{log(L/L$_{\odot}$)} & \colhead{T$_{\rm eff}$ (K)}  
}
\startdata
L0   &  11.62  &  10.85  &  10.33  & 0.77 &    1.29 & 13.54 & $-$3.52 & 2510 \\
L0.5 &  11.81  &  11.02  &  10.49  & 0.79 &    1.32 & 13.73 & $-$3.60 & 2400 \\
L1   &  12.00  &  11.19  &  10.65  & 0.81 &    1.35 & 13.92 & $-$3.67 & 2300 \\
L1.5 &  12.19  &  11.36  &  10.81  & 0.83 &    1.38 & 14.10 & $-$3.74 & 2200 \\
L2   &  12.38  &  11.54  &  10.98  & 0.84 &    1.40 & 14.29 & $-$3.82 & 2110 \\
L2.5 &  12.57  &  11.71  &  11.14  & 0.86 &    1.43 & 14.46 & $-$3.89 & 2030 \\
L3   &  12.76  &  11.88  &  11.30  & 0.88 &    1.46 & 14.63 & $-$3.96 & 1950 \\
L3.5 &  12.95  &  12.05  &  11.46  & 0.90 &    1.49 & 14.80 & $-$4.02 & 1870 \\
L4   &  13.14  &  12.22  &  11.62  & 0.92 &    1.52 & 14.97 & $-$4.09 & 1800 \\
L4.5 &  13.33  &  12.39  &  11.79  & 0.94 &    1.54 & 15.13 & $-$4.16 & 1740 \\
L5   &  13.52  &  12.57  &  11.95  & 0.95 &    1.57 & 15.29 & $-$4.22 & 1670 \\
L5.5 &  13.71  &  12.74  &  12.11  & 0.97 &    1.60 & 15.44 & $-$4.28 & 1620 \\
L6   &  13.90  &  12.91  &  12.27  & 0.99 &    1.63 & 15.58 & $-$4.34 & 1570 \\
L6.5 &  14.09  &  13.09  &  12.43  & 1.00 &    1.66 & 15.72 & $-$4.39 & 1520 \\
L7   &  14.28  &  13.26  &  12.59  & 1.02 &    1.69 & 15.85 & $-$4.44 & 1470 \\
L7.5 &  14.47  &  13.43  &  12.76  & 1.04 &    1.71 & 15.99 & $-$4.50 & 1430 \\
L8   &  14.66  &  13.60  &  12.92  & 1.06 &    1.74 & 16.11 & $-$4.55 & 1390 \\
T0.5 &  14.79  &  13.60  &  13.21  & 1.19 &    1.58 & 16.14 & $-$4.56 & 1380 \\
T1   &  14.60  &  13.58  &  13.23  & 1.02 &    1.37 & 16.10 & $-$4.54 & 1390 \\
T1.5 &  14.44  &  13.59  &  13.27  & 0.85 &    1.17 & 16.08 & $-$4.54 & 1400 \\
T2   &  14.33  &  13.63  &  13.35  & 0.70 &    0.98 & 16.09 & $-$4.54 & 1390 \\
T2.5 &  14.27  &  13.70  &  13.46  & 0.57 &    0.81 & 16.11 & $-$4.55 & 1390 \\
T3   &  14.26  &  13.80  &  13.60  & 0.46 &    0.66 & 16.20 & $-$4.58 & 1360 \\
T3.5 &  14.28  &  13.93  &  13.77  & 0.35 &    0.51 & 16.30 & $-$4.62 & 1330 \\
T4   &  14.36  &  14.10  &  13.96  & 0.26 &    0.40 & 16.42 & $-$4.67 & 1290 \\
T4.5 &  14.48  &  14.29  &  14.19  & 0.19 &    0.29 & 16.59 & $-$4.74 & 1240 \\
T5   &  14.64  &  14.52  &  14.45  & 0.12 &    0.19 & 16.78 & $-$4.82 & 1190 \\
T5.5 &  14.85  &  14.78  &  14.74  & 0.07 &    0.11 & 17.01 & $-$4.91 & 1130 \\
T6   &  15.11  &  15.07  &  15.06  & 0.04 &    0.05 & 17.27 & $-$5.01 & 1060 \\
T6.5 &  15.41  &  15.39  &  15.41  & 0.02 &    0.00 & 17.57 & $-$5.13 &  990 \\
T7   &  15.75  &  15.74  &  15.79  & 0.01 & $-$0.04 & 17.90 & $-$5.26 &  920 \\
T7.5 &  16.14  &  16.12  &  16.20  & 0.02 & $-$0.06 & 18.27 & $-$5.41 &  840 \\
T8   &  16.58  &  16.53  &  16.64  & 0.05 & $-$0.06 & 18.68 & $-$5.58 &  770 \\
\enddata
\end{deluxetable}


\begin{thebibliography}{}
\bibitem[Abazajian et al. 2003]{abazajian03}Abazajian, K., et al. 2003, \aj, 126, 2081
\bibitem[Basri 2000]{bas00} Basri, G. 2000, ARA\&A, 38, 485
\bibitem[Becklin \& Zuckerman 1988]{bec88} Becklin, E. E. \& Zuckerman, B. 1988,
Nature, 336, 656
\bibitem[Bessel \& Brett 1988]{bes88} Bessel, M. S. \& Brett, J. M. 1988, \pasp, 100, 1134
\bibitem[Burgasser et al. 1999]{bur99} Burgasser, A. J., et al. 1999, \apj, 522, L65
\bibitem[Burgasser et al. 2000]{bur00} Burgasser, A. J., et al. 2000, \apj, 531, L57 
\bibitem[Burgasser 2001]{bur01} Burgasser, A. J. 2001, Ph.D. Dissertation, California Institute
of Technology
\bibitem[Burgasser et al. 2002a]{bur02a} Burgasser, A. J., et al. 2002a, \apj, 564, 421
\bibitem[Burgasser et al. 2002b]{bur02b} Burgasser, A. J., Marley, M. S., Ackerman, A. S., Saumon, D.,
Lodders, K., Dahn, C. C., Harris, H. C., \& Kirkpatrick, J. D. 2002b, \apj, 571, L151
\bibitem[Burgasser et al. 2003a]{bur03a} Burgasser, A. J., Kirkpatrick, J. D., Reid, I. N.,
Brown, M. E., Miskey, C. L., \& Gizis, J. E. 2003a, \apj, 586, 512
\bibitem[Burgasser et al. 2003b]{bur03b} Burgasser, A. J., Kirkpatrick, J. D., Liebert, J., 
\& Burrows, A. 2003b, \apj, 594, 510
\bibitem[Burrows et al. 1997]{bur97} Burrows, A., et al. 1997, \apj, 491, 856
\bibitem[Carpenter 2001]{car01} Carpenter, J. M. 2001, \aj, 121, 2851
\bibitem[Chabrier \& Baraffe 2000]{cha00} Chabrier, G. \& Baraffe, I. 2000, ARA\&A, 38, 337
\bibitem[Cruz et al. 2003]{cru03} Cruz, K. L., Reid, I. N., Liebert, J., Kirkpatrick, J. D.,
\& Lowrance, P. J. 2003, \aj, 126, 2421
\bibitem[Dahn 1997]{dah97} Dahn, C. C. 1997, in IAU Symp. No. 189, Fundamental Stellar
Properties: The Interaction Between Observation and Theory, eds. T. R. Bedding, A. J. Booth,
\& J. Davis (Dordrecht: IAU), 19
\bibitem[D02]{dah02} Dahn, C. C., et al. 2002, \aj, 124, 1170 (D02)
\bibitem[Delfosse et al. 1997]{del97} Delfosse, X. et al. 1997, A\&A, 327, L25
\bibitem[Drilling \& Landolt 2000]{dri00} Drilling, J. S. \& Landolt, A. U. 2000, in Allen's
Astrophysical Quantities, Fourth Edition, ed. A. Cox (New York: Springer-Verlag), 382
\bibitem[Elias et al. 1982]{eli82} Elias, J. H., Frogel, J. A., Matthews, K.,
\& Neugebauer, G. 1982, \aj, 87, 1029 (erratum 87, 1893)
\bibitem[Enoch, Brown, \& Burgasser 2003]{eno03} Enoch, M. L., Brown, M. E., \& Burgasser, A. J. 2003, \aj, in press
\bibitem[Epchtein 1997]{epc97} Epchtein, N. 1997, in The Impact of Large Scale Near--IR Surveys,
ed. F. Garzon (Dordrecht: Kluwer), 15
\bibitem[Fan et al. 2000]{fan00} Fan, X. et al. 2000, \aj, 119, 928
\bibitem[Fischer et al. 2003]{fis03} Fischer, J., et al. 2003, in Proc. SPIE 4841, 
Instrument Design and Performance for Optical/Infrared Ground-based Telescopes, 
eds. M. Iye \& A. F. M. Moorwood (Bellingham: SPIE), 564 
\bibitem[Fowler et al. 1996]{fow96} Fowler, A. M., Gatley, I., McIntyre, P., Vrba, F. J.,
\& Hoffman, A. 1996, in Proc. SPIE 2816, Symposium on Infrared Detection for Remote
Sensing: Physics, Materials, and Devices, eds. R. E. Longshore \& J. W. Baars 
(Bellingham: SPIE), 150 
Shimasaku, K., \& Schneider, D. P. 1996, \aj, 111, 1748 
\bibitem[Geballe et al. 2001]{geb01} Geballe, T. R., Saumon, D., Leggett, S. K., Knapp, G. R.,
Marley, M. S., \& Lodders, K. 2001, \apj, 556, 373
\bibitem[Geballe et al. 2002]{geb02} Geballe, T. R., et al. 2002, \apj, 564, 466
\bibitem[Gizis et al. 2003]{giz03} Gizis, J. E., Reid, I. N., Knapp, G. R., Liebert, J., Kirkpatrick,
J. D., Koerner, D. W., \& Burgasser, A. J. 2003, \aj, 125, 3302
\bibitem[Golimowski et al. 2004]{gol04} Golimowski, D. A., et al. 2004, \aj, submitted
\bibitem[Guetter et al. 2003]{gue03} Guetter, H. H., Vrba, F. J., Henden, A. A.,
\& Luginbuhl, C. B. 2003, \aj, 125, in press
\bibitem[Hawarden et al. 2001]{haw01} Hawarden, T. G., Leggett, S. K., Letawsky, M. B.,
Ballantyne, D. R., \& Casali, M. M. 2001, \mnras, 325, 563 
\bibitem[Hawley et al. 2002]{haw02} Hawley, S. L., et al. 2002, \aj, 123, 3409
\bibitem[Kirkpatrick et al. 1999]{kir99} Kirkpatrick, J. D. et al. 1999. \apj, 519, 802
\bibitem[Kirkpatrick et al. 2000]{kir00} Kirkpatrick, J. D., et al. 2000, \aj, 120, 447
\bibitem[Kirkpatrick 2004]{kir04} Kirkpatrick, J. D., et al. 2004, in preparation
\bibitem[Knapp et al. 2004]{kna04} Knapp, G. R., et al. 2004, \aj, submitted
\bibitem[Koerner et al. 1999]{koe99} Koerner, D. W., Kirkpatrick, J. D., McElwain, M. W.,
\& Bonaventura, A. J. 1999, \apj, 526, L25
\bibitem[Leggett et al. 1999]{leg99} Leggett, S. K., Toomey, D. W., Geballe, T. R., \& 
Brown, R. H. 1999, \apj, 517, L139
\bibitem[Leggett et al. 2000]{leg00} Leggett, S. K. et al. 2000, \apj, 536, L35
\bibitem[Leggett et al. 2001]{leg01} Leggett, S. K., Allard, F., Geballe, T. R., 
Hauschildt, P. H., \& Schweitzer, A. 2001, \apj, 548, 908
\bibitem[Leggett et al. 2002]{leg02} Leggett, S. K. et al. 2002, \apj, 564, 452
\bibitem[Luginbuhl et al. 1998]{lug98} Luginbuhl, C. B., Henden, A. A., Vrba, F. J., \&
Guetter, H. H. 1998, in Proc. SPIE 3354, Infrared Astronomical Instrumentation, ed. A. Fowler
(Bellingham: SPIE), 240 
\bibitem[Luyten 1979]{luy79} Luyten, W. J. 1979, NLTT Catalog (University of Minnesota)
\bibitem[Mart\'{\i}n et al. 1999]{mar99} Mart\'{\i}n, E. L., Delfosse, X., Basri, G., 
Goldman, B., Forveille, T., \& Zapatero-Osorio, M. R. 1999, \aj, 118, 2466
\bibitem[Monet \& Dahn 1983]{mon83} Monet, D. G. \& Dahn, C. C. 1983, \aj, 88, 1489
\bibitem[Monet et al. 1992]{mon92} Monet, D. G., Dahn, C. C., Vrba, F. J., Harris, H. C.,
Pier, J. R., Luginbuhl, C. B., \& Ables, H. D. 1992, \aj, 103, 638
\bibitem[Nakajima et al. 1995]{nak95} Nakajima, T., Oppenheimer, B. R., Kulkarni, S. R.,
Golimowski, D. A., Matthews, K., Durrance, S. T. 1995, Nature, 378, 463
\bibitem[Oppenheimer et al. 1995]{opp95} Oppenheimer, B. R., Kulkarni, S. R., Matthews, K., \&
Nakajima, T. 1995, Science, 270, 1478
\bibitem[Reid et al. 2001]{rei01} Reid, I. N., Gizis, J. E., Kirkpatrick, J. D., \&
Koerner, D. W. 2001, \aj, 121, 489
\bibitem[Schlegel et al. 1998]{sch98} Schlegel, D. J., Finkbeiner, D. P., \& Davis, M. 1998, \apj, 500, 525
\bibitem[Schmidt-Kaler 1982]{sch82} Schmidt-Kaler, Th. 1982, Landolt-Borstein, New
Series, Group VI (New York: Springer), Subvolume 2b, 453
\bibitem[Schneider et al. 2002]{sch02} Schneider, D. P. et al. 2002, \aj, 123, 458
\bibitem[Siegel et al. 2002]{sie02} Siegel, M. H., Majewski, S. R., Reid, I. N., 
\& Thompson, I. B. 2002, \apj, 578, 151
\bibitem[Skrutskie et al. 1997]{skr97} Skrutskie, M. F. 1997, in The Impact of Large-Scale
Near--IR Sky Surveys, ed. F. Garzon et al. (Dordrecht: Kluwer), 25
\bibitem[Stephens 2003]{ste03} Stephens, D. C. 2003, in IAU Symp. 211, Brown Dwarfs, ed. E. Mart\'in
(San Francisco; ASP), 355
\bibitem[Stephens \& Leggett 2004]{ste04} Stephens, D. C. \& Leggett, S. K. 2004, \pasp, 116, 9
\bibitem[Stone 1997]{sto97} Stone, R. C. 1997, \aj, 114, 2811
\bibitem[Strauss et al. 1999]{str99} Strauss, M. A., et al. 1999, \apj, 522, L61
\bibitem[Tinney, Burgasser, \& Kirkpatrick 2003]{tin03} Tinney, C. G., Burgasser, A. J., \& Kirkpatrick,
J. D. 2003, \aj, 126, 975 (TBK03)
\bibitem[Tsuji \& Nakajima 2003]{tsu03} Tsuji, T. \& Nakajima, T 2003, \apj, 585, L151
\bibitem[Tsvetanov et al. 2000]{tsv00} Tsvetanov, Z. I. et al. 2000, \apj, 531, L61
\bibitem[Vrba et al. 2000]{vrb00} Vrba, F. J., Henden, A. A., Luginbuhl, C. B., Guetter, H. H.,
\& Monet, D. G. 2000, \baas, 32, 678
\bibitem[York et al. 2000]{york00} York, D. G., et al. 2000, \aj, 120, 1579 

\end{thebibliography}
\end{document}